\documentclass[journal=jacsat,manuscript=article]{achemso}

\usepackage[version=3]{mhchem} 

\usepackage{graphicx}
\usepackage{dcolumn}
\usepackage{bm}
\usepackage{mathrsfs}
\usepackage[]{xcolor}
\usepackage{siunitx}
\usepackage{float}
\usepackage{amsmath,amssymb}
\usepackage{booktabs}
\usepackage{multirow}
\usepackage{pdfpages}

\usepackage[normalem]{ulem}

\newcolumntype{C}[1]{>{\centering\let\newline\\\arraybackslash\hspace{0pt}}m{#1}}

\newcommand{\de}{\ensuremath{\, \mathrm{d}}}

\newcommand{\ket}[1]{\vert #1 \rangle}
\newcommand{\Ket}[1]{\Big\vert #1 \Big\rangle}
\newcommand{\bra}[1]{\langle #1 \vert}
\newcommand{\Bra}[1]{\Big\langle #1 \Big\vert}
\newcommand{\braket}[1]{\langle #1 \rangle}

\newcommand{\transp}{^\mathrm{T}}
\renewcommand{\mathbf}{\boldsymbol}

\title{Analytical evaluation of ground state gradients in quantum electrodynamics coupled cluster theory}

\author{Marcus T. Lexander}
\affiliation{%
 Department of Chemistry, Norwegian University of Science and Technology, 7491 Trondheim,
Norway
}
\alsoaffiliation{These authors contributed equally to this work}
\author{Sara Angelico}
\affiliation{%
 Department of Chemistry, Norwegian University of Science and Technology, 7491 Trondheim,
Norway
}
\alsoaffiliation{These authors contributed equally to this work}
\author{Eirik F.~Kjønstad}
\affiliation{%
 Department of Chemistry, Norwegian University of Science and Technology, 7491 Trondheim,
Norway
}%
\author{Henrik Koch}
\email{henrik.koch@ntnu.no}
\affiliation{%
 Department of Chemistry, Norwegian University of Science and Technology, 7491 Trondheim,
Norway
}

\date{\today}

\begin{document}

\begin{abstract}
\noindent
Analytical gradients of potential energy surfaces play a central role in quantum chemistry, allowing for molecular geometry optimizations and molecular dynamics simulations. In strong coupling conditions, potential energy surfaces can account for strong interactions between matter and the quantized electromagnetic field. In this paper, we derive expressions for the ground state analytical gradients in quantum electrodynamics coupled cluster theory. We also present a Cholesky-based implementation for the coupled cluster singles and doubles model. We report timings to show the performance of the implementation and present optimized geometries to highlight cavity-induced molecular orientation effects in strong coupling conditions. 
\end{abstract}
\maketitle
\section{1. Introduction}
In the Born-Oppenheimer approximation, nuclei evolve on electronic potential energy surfaces, driven by the force given by the gradient of the electronic energy.
The identification of relevant geometries on potential energy surfaces, as well as the study of chemical reactivity or orientational effects through molecular dynamics simulations, relies on the evaluation of molecular gradients of the potential energy surfaces. 
\vspace*{0.2cm}\\
An efficient evaluation of these gradients usually relies on an implementation of the analytical nuclear derivative of the electronic energy.\cite{SCHAEFER1986369,pulay2014analytical}  
Among several electronic structure methods, coupled cluster theory is well-known to provide a highly accurate description of dynamical correlation, both for ground and excited states, when the ground state is dominated by a single reference configuration.\cite{bartlett2007coupled,helgaker2013molecular} 
In addition, it is known to converge rapidly to the exact limit as one moves up the hierarchy of methods.\cite{loos2018mountaineering,loos2020mountaineering} 
Due to its increasingly feasible computational cost, its singles and doubles formulation (CCSD)\cite{purvis1982full} is today extensively used for calculations of energies and different properties for medium sized systems. Many implementations of analytical gradients at the CCSD level have been reported over the past decades.\citep{feng2019implementation,kallay2003analytic,gauss1991coupled,hald2003lagrangian,bozkaya2016analytic,Schnack2022efficient} More recently, decomposition methods for the electronic repulsion integrals have been used to further improve the efficiency of such gradient algorithms.\cite{Schnack2022efficient,feng2019implementation,delcey2014analytical,bozkaya2016analytic} 
In particular, the Cholesky decomposition method, which provides rigorous error thresholds, have recently become applicable to much larger systems due to algorithmic advances.\cite{aquilante2011cholesky,folkestad2019efficient,pedersen2024versatility}
\vspace*{0.2cm}\\
In strong coupling conditions, the strong interactions between light and matter lead to the formation of hybrid light-matter states named polaritons. In these conditions, several experimental studies have shown modifications of e.g. ground state chemical and photochemical reactivity\cite{thomasgroundstate,saumodifying,ahn2023modification,hutchison2012modifying,thomas2019tilting} and supramolecular organization.\cite{biswas2024electronic,joseph2021supramolecular,hirai2021selective,sandeep2022manipulating,joseph2024consequences} A rationalization of such modifications, however, requires a detailed description of the quantum nature of both the molecule and the electromagnetic field. In this direction, many quantum chemistry \textit{ab initio} methods have been generalized to quantum electrodynamics (QED).
Recent examples include QED density functional theory,\cite{ruggenthaler2014quantum,pellegrini2015optimized,lu2024electron,vu2022enhanced} QED Hartree-Fock,\cite{Haugland2020Coupled,deprince2021cavity} QED configuration interaction,\cite{Haugland2020Coupled,vu2024cavity,mctague2022non} and QED coupled cluster theory .\cite{Haugland2020Coupled,mordovina2020polaritonic,pavosevic2021polaritonic}
While the \textit{ab initio} character of these methods provides a proper description of the molecular system, an accurate treatment of cavity-mediated reorientation effects, as well as changes in the equilibrium geometry, is needed in order to make more robust predictions.\cite{castagnola2024polaritonic,liebenthal2024orientation,schnappinger2024molecular} 
To this end, implementations of analytical gradients of some \textit{ab initio} methods have already been implemented.\cite{liebenthal2024orientation,schnappinger2023abinitio,yang2022cavity} 
\vspace*{0.2cm}\\
In this work, we also move towards this end-goal, presenting a general formulation of analytical gradients for the ground state energy in QED coupled cluster (QED-CC) theory, along with an implementation at the QED-CC with single and doubles electronic excitations and single photonic excitations (QED-CCSD-1) level.
We provide timings of the {QED-CCSD-1} gradient evaluations in order to 
demonstrate the efficiency of the implementation, as well as optimized geometries in various systems to highlight the importance of cavity-induced orientation and relaxation effects.
\section{2. Theory}
\subsection{2A. QED Hamiltonian}
\noindent
In strong coupling conditions, the Hamiltonian must include the quantized electromagnetic field and its interactions with matter. 
Here, we describe such a system by means of the Pauli-Fierz Hamiltonian expressed within the length gauge representation of the dipole approximation.\cite{ruggenthaler2023understanding, castagnola2024polaritonic} Moreover, we adopt the Born-Oppenheimer approximation, assuming that the wave function for the electronic and photonic degrees of freedom depends only on the positions of the nuclei. Working in the QED-HF coherent-state basis for a single mode of the electromagnetic field, we finally obtain the electronic-photonic Hamiltonian\cite{Haugland2020Coupled,foley2023ab}
\begin{equation}
\label{eq-PF Hamiltonian}
    H = H_e + \omega b^\dagger b + \sqrt{\frac{\omega}{2}}    (\boldsymbol{\lambda}\cdot(\mathbf{d}-\langle\mathbf{d}\rangle))(b^\dagger+b)+   \frac12(\boldsymbol{\lambda}\cdot(\mathbf{d}-   \langle\mathbf{d}\rangle))^2,
\end{equation}
where $\mathbf{d}$ is the dipole moment operator and $\langle \mathbf{d}\rangle$ is its expectation value at the QED Hartree-Fock (QED-HF) level. The electromagnetic field is represented by a single harmonic oscillator with frequency $\omega$ and photon creation and annihilation operators denoted by $b^\dagger$ and $b$. \\
The first two terms of eq.~\eqref{eq-PF Hamiltonian} describe the molecular electronic Hamiltonian, $H_e$, and the energy of the quantized electromagnetic field, respectively. 
The third term in $H$ describes the bilinear interaction between the molecular system and the displacement field. 
The final term is the dipole self-energy, which ensures that the Hamiltonian is bounded from below.\cite{schafer2020relevance,rokaj2018light} The coupling strength of the field is denoted by $\boldsymbol{\lambda} = \sqrt{\frac{4\pi}{V}}\boldsymbol{\varepsilon}$, where $V$ is the quantization volume and $\boldsymbol{\varepsilon}$ the polarization vector. To simplify the notation, we will let $d = \boldsymbol{\lambda}\cdot\mathbf{d}$ in the following. \\
By expanding the electronic Hamiltonian in the second quantization formalism, eq.~\eqref{eq-PF Hamiltonian} can be rewritten as
\begin{equation}
\begin{split}
\label{eq:2nd-quant-PF-Ham}
    H &= \sum_{pq} h_{pq}E_{pq} + \frac12\sum_{pqrs}g_{pqrs}e_{pqrs} + \omega b^\dagger b \\
    &+ \sqrt{\frac{\omega}{2}} \sum_{pq}d_{pq}E_{pq}(b^\dagger + b) - \sqrt{\frac{\omega}{2}} \langle {d}\rangle (b^\dagger + b) + h_{nuc} .
\end{split}
\end{equation}
Here $h_{nuc}$ is the nuclear repulsion energy. The indices $p,q,r,s$ denote molecular orbitals (MOs), and 
\begin{equation}
    E_{pq} = \sum_\sigma a^\dagger_{p\sigma}a_{q\sigma} \hspace{40pt} e_{pqrs} = E_{pq}E_{rs} - \delta_{qr}E_{ps} , 
\end{equation}
where $a^\dagger$ and $a$ are the creation and annihilation operators for the electrons, respectively. 
Moreover, the one- and two-electron integrals ${h}_{pq}$ and ${g}_{pqrs}$ are dressed electronic integrals that include contributions from the electromagnetic field. Denoting the electronic one- and two-electron integrals as $h_{pq}^e$ and $g_{pqrs}^e$, we have
\begin{align}
    &h_{pq} = h^e_{pq} + \frac12 \sum_r {d}_{pr} {d}_{rq} - {d}_{pq}\langle{d}\rangle + \frac{\delta_{pq}}{2N_e} \langle {d}\rangle^2 \\
    &g_{pqrs} = g^e_{pqrs} + {d}_{pq}{d}_{rs},
\end{align}
where $N_e$ is the number of electrons of the molecule, and $d_{pq}$ can be split in an electronic part $d_{pq}^e$ and a nuclear part $d_N$: 
\begin{equation}
d_{pq} = d_{pq}^e + \frac{S_{pq}}{N_e}d_N.
\end{equation}
Note that, when introducing the second quantization formalism for the electronic Hamiltonian, we have implicitly made use of the complete basis set approximation (i.e., we have approximated the square of the second quantization dipole moment operator $\hat{d}^2 = \hat{d}\cdot\hat{d}$).\cite{foley2023ab} The generalization of the current implementation explicitly including the quadrupole moment is straightforward.
\subsection{2B. QED-CC}
\noindent
In QED-CC, coupled cluster theory is extended to include the interactions between the electrons and the quantized electromagnetic field.
The wave function is obtained by applying the exponential of the cluster operator $T$ to a reference wave function, which is typically chosen to be the QED-HF wave function:\cite{Haugland2020Coupled}
\begin{equation}
    \ket{\text{QED-CC}} = e^T \ket{\text{QED-HF}} = e^T\ket{\text{HF,0}}.
    \label{eq:wavefunction}
\end{equation}
The cluster operator is defined as
\begin{equation}
    T = \sum_{\mu, n \in \mathscr{E}}{\zeta_{\mu n} \tau_\mu (b^\dagger)^n}, 
\end{equation}
where
\begin{align}
\begin{split}
    \mathscr{E} = \big\{ &(\mu,n) : \mu = 0, \; n > 0, \\ &(\mu,n) : \mu > 0, \; n \geq 0 \big\}
\end{split}
\end{align}
denotes the set of excitation operators  in $T$. In particular, $\mathscr{E}$ contains all elements in the projection manifold except $\ket{\mathrm{HF},0}$. Here, $\mu$ labels the electronic excitations, with $\mu = 0$ denoting the HF state, and $n$ denoting the photonic excitation.
We can partition the cluster operator into purely electronic, purely photonic, and mixed excitation operators,
\begin{equation}
  T = T_e + T_p + T_{int},
\end{equation}
with
\begin{equation}
    T_e = \sum_{\mu \geq 1} t_{\mu}\tau_{\mu} \hspace{40pt}
    T_p = \sum_{n \geq 1}{\gamma_n (b^\dagger)^n} \hspace{40pt}
    T_{int} = \sum_{\substack{\mu \geq 1 \\ n \geq 1}} s^n_{\mu} \tau_{\mu} (b^\dagger)^n
\end{equation}
Moreover, we will refer to $\zeta_{\mu n} = \{t_\mu, \gamma_n, s_\mu^n\}$ as the QED-CC amplitudes.
We can determine the QED-CC energy and amplitudes by projecting the Schrödinger equation on the excitation set:\cite{Haugland2020Coupled}
\begin{align}
    \label{eq:cc-energy}
    E &= \bra{\text{HF},0}\bar{H}\ket{\text{HF},0} \\
    \label{eq:cc-omega-mu-n}
    \Omega_{\mu n} &= \bra{\mu, n} \bar{H}\ket{\text{HF},0} = 0, \quad (\mu, n) \in \mathscr{E},
\end{align}
where $\bar{H}=e^{-T}He^T$ is the similarity transformed Hamiltonian. \\
While these equations provide expressions for the energy and the amplitudes, the direct evaluation of the molecular gradient as total derivative of the energy of eq.~\eqref{eq:cc-energy} is complicated and usually avoided. For this reason, in the next section, we describe the Lagrangian formalism, which is commonly used to derive analytical expressions for the molecular gradients in coupled cluster theory and other electronic structure methods.\cite{stanton1993many,koch1990coupled,Handy1984evaluation,helgaker1992calculation} 
\subsection{2C. The Lagrangian formalism}
\label{Sec:Lagrangian-formalism}
\noindent
In coupled cluster theory, the dependence of the ground state energy on the amplitudes is non-variational; that is, the energy is not stationary with respect to the amplitudes.
As a consequence, we cannot invoke the usual Hellmann-Feynman theorem to calculate nuclear gradients. Nevertheless, we can avoid explicitly evaluating the derivatives of the wave function parameters by adopting the Lagrangian formalism (or Z-vector technique).\cite{Handy1984evaluation,helgaker1992calculation} 
\vspace*{0.2cm}\\
In this formalism, we consider a function (for example, the energy $E$) that depends on some parameters $\boldsymbol{\lambda}$. 
These parameters are determined by imposing a set of conditions \{$e_p=0\}$. 
To each of these constraints $e_p$, we can now associate a Lagrangian multiplier $\bar{\lambda}_p$ and define a Lagrangian $\mathcal{L}$ as
\begin{equation}
    \mathcal{L}(\boldsymbol{\lambda},\boldsymbol{\bar{\lambda}},\mathbf{x}) = E(\boldsymbol{\lambda},\mathbf{x})+ \boldsymbol{\bar{\lambda}}\transp\mathbf{e}(\boldsymbol{\lambda},\mathbf{x}) ,
\end{equation}
where $\mathbf{x}$ denotes the nuclear coordinates. In order to keep the notation simple, we will not make the dependence on $\mathbf{x}$ explicit in the following. \\
To enforce the set of constraints $\mathbf{e}=0$ and determine the parameters $\boldsymbol{\lambda}$, we require that the Lagrangian is stationary with respect to the multipliers $\boldsymbol{\bar{\lambda}}$. 
The multipliers, instead, are determined by requiring that $\mathcal{L}$ is stationary with respect to the parameters $\boldsymbol{\lambda}$:
\begin{equation}
   \frac{\partial \mathcal{L}}{\partial \mathbf{\bar{\lambda}}} = 0 \Rightarrow \mathbf{e}=0, \quad  \frac{\partial \mathcal{L}}{\partial \mathbf{\lambda}} = 0 .
\end{equation}
When these equations are satisfied, the Lagrangian is by definition equal to the energy at every value of $\boldsymbol{x}$. 
As a consequence, the total derivative of the energy with respect to $\boldsymbol{x}$ can be evaluated as the partial derivative of the Lagrangian:
\begin{equation}
\label{eq:nuclear-derivative-in-lagr-formalism}
    \frac{\de E}{\de x} = \frac{\de \mathcal{L}}{\de x} = \frac{\partial \mathcal{L}}{\partial x} + \sum_p \frac{\partial \mathcal{L}}{\partial \lambda_p} \frac{\partial \lambda_p}{\partial x}+\sum_p \frac{\partial \mathcal{L}}{\partial \bar{\lambda}_p} \frac{\partial \bar{\lambda}_p}{\partial x} = \frac{\partial \mathcal{L}}{\partial x}.
\end{equation}
In this way, the molecular gradient can be evaluated without needing to evaluate the derivative of the parameters with respect to the nuclear coordinates. \\
Applying this formalism to the QED-CC ground state energy, we can define the Lagrangian as
\begin{equation}
\label{eq:Lagrangian-explicit}
    \mathcal{L} = \bra{\text{HF},0}e^{-T}e^{\kappa}He^{-\kappa}e^T\ket{\text{HF},0} + \sum_{\mu, n \in \mathscr{E}} \bar{\zeta}_{\mu n} \bra{\mu, n}e^{-T}e^{\kappa}He^{-\kappa}e^T\ket{\text{HF},0} + \sum_{ai}\Bar{\kappa}_{ai}F_{ai} .
\end{equation}
Here, the first term is the QED-CC ground state energy, and the second term corresponds to the $\Omega_{\mu n}$ equations and the associated multipliers $\bar{\zeta}_{\mu n}$. The last term enforces the QED-HF equations through the associated multipliers $\bar{\kappa}_{ai}$. Moreover, we have made the $\boldsymbol{x}$-dependence of the orbitals explicit by introducing the orbital rotation operator $\kappa$. By assumption, the QED-HF orbitals are optimized, and $\kappa=0$, at the geometry where the gradient is evaluated.\cite{helgaker1992calculation}   
Finally, we can rewrite the QED-CC Lagrangian in a more compact form by introducing the dual ground state vector $\bra{\Lambda}$:
\begin{align}
    &\bra{\Lambda} = \Big(\bra{\text{HF},0} + \sum_{\mu,n \in \mathscr{E}}\bar{\zeta}_{\mu n}\bra{\mu,n}\Big)e^{-T}\\
    \label{eq:Lambda-lagrangian}
   \mathcal{L} = \bra{\Lambda} e^{\kappa} H e^{-\kappa}&\ket{\text{QED-CC}} + \sum_{ai}\bar{\kappa}_{ai}F_{ai} =
   \bra{\Lambda}\tilde{H}\ket{\text{QED-CC}} + \sum_{ai}\bar{\kappa}_{ai}F_{ai} ,
\end{align}
where we have defined $\tilde{H}= e^\kappa H e^{-\kappa}$.
\subsection{2D. Gradient expression}
\noindent 
Using eqs.~\eqref{eq:nuclear-derivative-in-lagr-formalism} and~\eqref{eq:Lambda-lagrangian}, the molecular gradient can be expressed as
\begin{equation}
    \frac{\de E}{\de x} = \mathcal{L}^{(1)} = \bra{\Lambda}\tilde{H}^{(1)}\ket{\text{QED-CC}} + \sum_{ai}\bar{\kappa}_{ai}F_{ai}^{(1)},
\end{equation}
where we have denoted nuclear derivatives with $^{(1)}$. \\
So far, we have taken care of the constraints for the QED-HF and QED-CC equations, which are imposed via the Lagrangian. However, when evaluating molecular gradients, we must also ensure that the MOs are kept orthonormal at all nuclear geometries, since we  implicitly assume this in all our derivations. In fact, the Hamiltonian and other operators, as well as the state vectors, are represented in terms of creation and annihilation operators ($a_{p\sigma}^\dagger$, $a_{p\sigma}$) that are associated with a set of orthonormal MOs ($\phi_p$).\citep{helgaker2013molecular}
To account for orthonormality, we employ an orbital connection. 
Given a reference geometry $\boldsymbol{x}_0$, at which we will calculate the gradient, we can consider some unmodified MOs (UMOs) at a neighbouring geometry $\boldsymbol{x}$,
\begin{align}
    \phi_p^\mathrm{UMO}(\boldsymbol{x}) = \sum_{\alpha} C_{\alpha p}(\boldsymbol{x}_0) \chi_{\alpha}(\boldsymbol{x}),
\end{align}
formed by freezing the orbital coefficients $C_{\alpha p}$ at $\boldsymbol{x}_0$. These UMOs are not orthonormal, and we denote the overlap matrix as $S_{pq}(\boldsymbol{x})=\braket{\phi_p^\mathrm{UMO}(\boldsymbol{x}) |\phi_q^\mathrm{UMO}(\boldsymbol{x})}$.
An orbital connection $\boldsymbol{T}$ restores orthonormality by transforming the UMOs into a set of orthonormal MOs (OMOs),
\begin{equation}
\label{eq:OMO}
    \phi^{\mathrm{OMO}}_{p} = \sum_q T_{pq}(\boldsymbol{x})  \phi_q^\mathrm{UMO}(\boldsymbol{x}).
\end{equation}
In this paper, we adopt the symmetric connection $\boldsymbol{T} = \boldsymbol{S}^{-1/2}$.\cite{olsen1995orbital} \\
From eq.~\eqref{eq:OMO}, it follows that we can separate the derivatives of the Hamiltonian into two contributions. The first one arises from the UMOs and the second one from the $\boldsymbol{x}$-dependence of $\boldsymbol{T}$. To evaluate the latter, we note that
 $   \boldsymbol{T}^\dagger \boldsymbol{S} \boldsymbol{T} = \boldsymbol{T} \boldsymbol{S} \boldsymbol{T} = \boldsymbol{1}$,
and take the derivative at $\boldsymbol{x}_0$ (where $\boldsymbol{S} = \boldsymbol{1}$). We then find
\begin{equation}
    2 \boldsymbol{T}^{(1)} + \boldsymbol{S}^{[1]} = 0 \implies \boldsymbol{T}^{(1)} = -\frac{1}{2} \boldsymbol{S}^{[1]}.
\end{equation}
The $^{[1]}$ notation denotes that the derivative is taken in the UMO basis.
Finally, we can consider the 
one-electron part of the Hamiltonian, $h$. We can write the $h$ derivative at $\boldsymbol{x}_0$ as
\begin{align}
    h^{(1)} = \sum_{pq} h^{(1)}_{pq} E_{pq} = \sum_{pq} h^{[1]}_{pq} E_{pq} + \sum_{pq} \sum_{rs} (T_{pr}^{(1)\dagger} T_{qs} + T_{pr}^\dagger T_{qs}^{(1)}) h_{rs} E_{pq}.
\end{align}
Now, since $T_{pq}(\boldsymbol{x}_0) = \delta_{pq}$, we can simplify this expression to 
\begin{align}
    h^{(1)} = \sum_{pq} h^{[1]}_{pq} E_{pq} - \frac{1}{2} \sum_{pq} \Bigl( \sum_{r} S^{[1]}_{rp} h_{qr} + \sum_s S^{[1]}_{qs} h_{ps} \Bigr) E_{pq}.
\end{align}
Expressing the two one-index transformations by $\{ ... \}$, we find that
\begin{align}
    h^{(1)} = h^{[1]} - \frac{1}{2} \{ S^{[1]}, h \}.
\end{align}
In the case of the two-electron part of the Hamiltonian, $g$, we obtain four one-index transformations between $S^{[1]}$ and $g$ when taking the derivative. Collecting the one- and two-electron terms with the notation $\{ ... \}$, we can write the derivative of the Hamiltonian operator as
\begin{align}
\label{eq:reortho-derivative-Hamiltonian}
    H^{(1)} = H^{[1]} - \frac{1}{2} \{ S^{[1]}, H \}.
\end{align}
Terms arising from the one-index transformations are referred to as ``reorthonormalization terms'' and will be considered in more detail below. In the above, we have used that the creation and annihilation operators can be considered independent of $\boldsymbol{x}$ in the case of energy derivatives.\citep{olsen1995orbital}
\vspace*{0.2cm}\\
The final expression for the gradient reads
\begin{equation}
\begin{split}
\label{eq:nuclear-gradient-densities}
    \frac{\de E}{\de x} \bigg{|}_0 &= \sum_{pq}{h}_{pq}^{(1)} D_{pq}^{e} + \sum_{pqrs} {g}^{(1)}_{pqrs}d_{pqrs}^{e} + h_{nuc}^{(1)} \\
    &+ \sqrt{\frac{\omega}{2}}\sum_{pq}d_{pq}^{(1)} D_{pq}^{e\text{-}p} - \sqrt{\frac{\omega}{2}}\langle d\rangle^{(1)}D^{p} + \sum_{ai}\Bar{\kappa}_{ai}F_{ai}^{(1)}
\end{split}
\end{equation}
where we have introduced the densities
\begin{align}
   &D_{pq}^e = \bra{\Lambda}E_{pq}\ket{\text{QED-CC}}   \\
    &d_{pqrs}^e = \bra{\Lambda}e_{pqrs}\ket{\text{QED-CC}} \\
   &D_{pq}^{e\text{-}p} = \bra{\Lambda}E_{pq}(b^\dagger+b)\ket{\text{QED-CC}} \\   
   &D^{p} = \bra{\Lambda}(b^\dagger+b)\ket{\text{QED-CC}}. 
\end{align}
The molecular gradient thus depends on electronic, photonic, and mixed electronic-photonic densities. 
Autogenerated programmable expressions are given in the Supporting Information. 
\subsection{2E. Response equations}
\noindent
In order to evaluate the nuclear gradient, we first need to determine the Lagrangian multipliers by solving two sets of response equations. 
First, by considering the derivative of the Lagrangian with respect to the QED-CC amplitudes, one gets the response (or stationarity) equation for the QED-CC multipliers $\boldsymbol{\bar{\zeta}}$:
\begin{equation}
    \boldsymbol{\bar{\zeta}}\boldsymbol{A} = -\boldsymbol{\eta} .
    \label{eq:multipliers}
\end{equation}
Here, the Jacobian $\boldsymbol{A}$ and the $\boldsymbol{\eta}$ vectors are the analogs of the standard equation of motion coupled cluster quantities,\cite{Haugland2020Coupled}
\begin{align}
  &  A_{\mu n,\nu m} = \bra{\mu, n}[ \bar{H},\tau_\nu(b^\dagger)^m]\ket{\text{QED-HF}} \\
  &  \eta_{\mu n} = \bra{\text{QED-HF}} [\bar{H},\tau_\mu(b^\dagger)^n]\ket{\text{QED-HF}} .
\end{align}
The derivative of the Lagrangian with respect to the orbital rotation parameters, instead, gives the response equation for the $\boldsymbol{\bar{\kappa}}$ multipliers:
\begin{equation}
    \label{eq:kappa-bar}
\boldsymbol{\Bar{\kappa}}\boldsymbol{A}^{\boldsymbol{\bar{\kappa}}} = -  \boldsymbol{\eta}^{\boldsymbol{\bar{\kappa}}}.
\end{equation}
Here, $\boldsymbol{A}^{\boldsymbol{\bar{\kappa}}}$ is the QED-HF Hessian,
\begin{equation}
    A^{\boldsymbol{\bar{\kappa}}}_{ai,bj} = \frac{\partial F_{ai}}{\partial \kappa_{bj}} = \delta_{ab}\delta_{ij}(\epsilon_a - \epsilon_i) + 4 g_{aibj} - g_{abij} - g_{ajbi} - 4d_{ai}d_{bj}, 
\end{equation}
while $\boldsymbol{\eta}^{\boldsymbol{\bar{\kappa}}}$ can be expressed as
\begin{equation}
\begin{split}
\label{eq:eta-hartree-fock}
    {\eta}^{\boldsymbol{\bar{\kappa}}}_{ai}= \Bra{\Lambda}\frac{\partial \tilde{H}}{\partial \kappa_{ai}}\Ket{\text{QED-CC}} &= \sum_{r} {h}_{ri}(D_{ra}^{e}+D_{ar}^{e}) - \sum_r {h}_{ra}(D_{ri}^{e} + D_{ir}^{e})  \\
    & +\sum_{rst} \big({g}_{irst}(d^e_{arst}+d^e_{rast}) - {g}_{arst}(d^e_{rist}+d^e_{irst})\big) \\
    & + \sqrt{\frac{\omega}{2}}\sum_r d_{ri}(D_{ra}^{e\text{-}p} + D_{ar}^{e\text{-}p}) - \sqrt{\frac{\omega}{2}}\sum_r d_{ra}(D_{ri}^{e\text{-}p} + D_{ir}^{e\text{-}p}) \\
    &+ 4d_{ai}\bigg(\sum_{rs}d_{rs}D_{rs}^{e} -\langle d\rangle +\sqrt{\frac{\omega}{2}}D^{p} \bigg) .
\end{split}
\end{equation}
Here, we have used 
\begin{equation}
    \frac{\partial \langle d \rangle}{\partial \kappa_{pq}} = 4d_{pq}(v_p - v_q) 
\end{equation}
with $v_p = 1$ if $p$ denotes an occupied orbital and $v_p = 0$ otherwise. All partial derivatives are evaluated at $\boldsymbol{\kappa}=0$. Note that while in the first two lines of eq.~\eqref{eq:eta-hartree-fock} the standard definition\cite{hald2003lagrangian} of the $\boldsymbol{\eta}^{\boldsymbol{\bar{\kappa}}}$ vector is obtained (albeit in terms of the dressed one- and two-electron integrals), new contributions due to the quantized electromagnetic field arise in the last two lines.
\section{3. Implementation}
In the following, we will work at the QED-CCSD-1 level of theory, where $T_e$ includes single and double electronic excitations and $T_p$ includes single photonic excitations.  
Additionally, $T_{int}$ includes simultaneous electron-photon excitations obtained by combining the included electronic and photonic excitations:
\begin{equation}
T_e = T_1 + T_1 \qquad T_p = \Gamma^ 1 \qquad T_{int} = S_1^1 + S_2^1
\end{equation}
where
\begin{equation}
\begin{split}
 &T_1 = \sum_{ai} t_{ai}E_{ai} \hspace{50pt}   T_2 = \frac12 \sum_{aibj} t_{aibj} E_{ai}E_{bj} \\
 &\Gamma^1 = \gamma b^\dagger \\
 &S_1^1 = \sum_{ai} s_{ai}E_{ai}b^\dagger   \hspace{38pt}
S_2^1 = \frac12 \sum_{aibj} s_{aibj} E_{ai}E_{bj}b^\dagger.
\end{split}
\end{equation}
Here, $i, j$ denote occupied orbitals and $a, b$ denote virtual orbitals. To determine the ground state energy and amplitudes, we solve eq.~\eqref{eq:cc-omega-mu-n} using the  projection set
\begin{equation}
    \{ \ket{\text{HF},0}, \ket{\mu,0}, \ket{\text{HF},1}, \ket{\mu,1}\},
\end{equation}
where $\mu$ is restricted to single and double excitations. \vspace*{0.2cm}\\
\noindent
Our implementation of QED-CCSD-1 gradients builds on the EOM-CCSD nuclear gradient implementation by Schnack-Petersen et al.,\citep{Schnack2022efficient} which makes use of Cholesky-decomposed two-electron integrals to evaluate the gradient. Here, we  highlight the aspects of this implementation that are most relevant to the present work and refer to Ref.~\citenum{Schnack2022efficient} for more details.\\
The two-electron integral matrix is sparse and positive definite and admits to a Cholesky decomposition, which can be expressed directly or in a resolution-of-identity form,\citep{beebe1977simplifications}
\begin{equation}
\label{eq:Cholesky-resolution-identity}
    g_{\alpha\beta\gamma\delta} = \sum_J L_{\alpha\beta}^J L_{\gamma\delta}^J = \sum_{JK} (\alpha \beta \vert J) (J \vert K)^{-1} (K \vert \gamma \delta),
\end{equation}
where the $J$ and $K$ indices denote AO index pairs that are referred to as Cholesky pivots and $\alpha,\beta,\gamma,\delta$ denote AO indices. Using the resolution of identity form, we can write the contribution of the two-electron terms in the energy (in the MO basis) as:
\begin{equation}
    \frac12 \sum_{pqrs}d^e_{pqrs}{g}_{pqrs} = \frac12 \sum_{pqrs} d^e_{pqrs}\sum_{JK}{(pq|J)}{(J|K)}^{-1}{(K|rs)} = \frac12\sum_{pqJ} L_{pq}^J \tilde{W}_{pq}^J
\end{equation}
where we have defined
\begin{equation}
    \tilde{W}_{pq}^{J} = \sum_{rs}d^e_{pqrs}L^{J}_{rs} .
\end{equation}
The resolution-of-identity form is particularly useful because it allows us to evaluate nuclear derivatives without determining the derivatives of the Cholesky vectors ($L^J_{\alpha\beta}$).\citep{delcey2014analytical,Schnack2022efficient} Moreover, the introduction of the intermediate $\tilde{W}_{pq}^J$ allows us to avoid storing the memory-intensive four-index block of the density matrix (e.g., $d_{abcd}$), whose contributions are instead stored in the three-index tensor $\tilde{W}_{pq}^J$.\cite{Schnack2022efficient} 
\subsection{3A. Response equations}
\noindent
In the case of the amplitude response, $\boldsymbol{A}$ and $\boldsymbol{\eta}$ have already been implemented, and we refer the reader to Ref.~\citenum{Haugland2020Coupled} for more details. In the case of the orbital relaxation, 
$\bar{\kappa}_{ai}$ is determined by solving eq.~\eqref{eq:kappa-bar}. The implementation of the right-hand side makes use of the $\tilde{W}^J_{pq}$ intermediate and the permutation operator $P_{ai}$ ($P_{ai}X_{ai} = X_{ia}$):
\begin{equation}
\begin{split}
     {\eta}^{\boldsymbol{\bar{\kappa}}}_{ai} &= (1-P_{ai})\bigg(\sum_{r} {h}_{ri}\mathcal{D}_{ra}^{e} 
     +\sum_{tJ} \tilde{\mathcal{W}}_{ti}^J L_{at}^J
    + \sqrt{\frac{\omega}{2}}\sum_r d_{ri}\mathcal{D}_{ra}^{e\text{-}p} \bigg) \\
    &+ 4d_{ai}\bigg(\sum_{rs}d_{rs}D_{rs}^{e} -\langle d\rangle +\sqrt{\frac{\omega}{2}}D^{p} \bigg)
\end{split}
\end{equation}
where we have introduced the symmetrized quantities
\begin{equation}
    \mathcal{D}_{pq}^X = D_{pq}^X + D_{qp}^X \hspace{40pt} \tilde{\mathcal{W}}_{pq}^J = \tilde{W}_{pq}^J + \tilde{W}_{qp}^J .
\end{equation}
Finally, the implementation of the QED-HF Hessian transformation, starting from an existing HF implementation,\cite{Schnack2022efficient}  is straightforward provided that the one- and two-electron integrals are properly redefined to include the QED contributions.
\subsection{3B. Nuclear gradient}
\noindent
Once the response equations have been solved, the nuclear gradient can be calculated. As mentioned before, this is usually split in two contributions. At first, UMOs contributions are considered. These include one- and two-electron contributions, both from the energy and the orbital relaxation terms, as well as contributions coming from the bilinear term of the Hamiltonian. The one-electron and bilinear contributions are straightforward and will not be discussed further. Below, we  describe the two-electron and reorthonormalization terms in more detail, emphasizing the required modifications to obtain the QED-CCSD-1 quantities starting from an existing CCSD implementation.
\subsubsection{Two-electron contributions}
Using the resolution-of-identity form in eq.~\eqref{eq:Cholesky-resolution-identity}, we find that the two-electron contributions to the gradient (in the UMO basis) can be written
\begin{align}
 \label{eq:2e-contraction-Cholesky-CCSD}
\begin{split}
    \frac{1}{2} \sum_{pqrs} d^e_{pqrs} g_{pqrs}^{[1]} &= \frac{1}{2} \sum_{pqrs} d^e_{pqrs} \Bigl( \sum_{JK} (pq \vert J) (J \vert K)^{-1} (K \vert rs) \Bigr)^{[1]} \\ 
    &= \sum_{pqJ} (pq \vert J)^{[1]} W_{pq}^J - \frac{1}{2} \sum_{ML} V_{ML} (M \vert L)^{(1)},
\end{split}
\end{align}
where we have defined the Cholesky intermediates
\begin{align}
    Z_{pq}^M &= \sum_J (pq \vert J) (J \vert M)^{-1}  \\ 
    \label{eq:W_pq^L}
    W_{pq}^L &= \sum_{rs} d^e_{pqrs} Z_{rs}^L = \tilde{W}_{pq}^{J}(J|L)^{-\frac12} \hspace{40pt}    V_{ML} = \sum_{pq} Z_{pq}^M W_{pq}^L.
\end{align}
A similar strategy can be used to evaluate the two-electron contributions to the orbital relaxation gradient. In this case, similar intermediates to $W_{pq}^J$ are defined from the contraction of $Z_{pq}^J$ with $\bar{\kappa}_{ai}$. A detailed description of these intermediates can be found in Ref.~\citenum{Schnack2022efficient}. 
Note that the $W_{pq}^L$ intermediates here introduced are calculated from $\tilde{W}_{pq}^L$. In this way, contractions with the two-electron density matrix only need to be evaluated once. 
\vspace*{0.2cm}\\
From the given list of the Cholesky pivots $\{ J \}$, we can evaluate the gradient by requesting $(\alpha\beta \vert J)^{(1)}$ and $(J \vert K)^{(1)}$ from an integral program, then transforming the $\alpha$ and $\beta$ indices to the MO basis, and, finally, evaluating the contractions in eq.~\eqref{eq:2e-contraction-Cholesky-CCSD} using the pre-calculated Cholesky intermediates. Hence, the memory-intensive four-index blocks of the density matrix do not need to be stored, as smaller batches of these densities can be constructed and immediately contracted with the Cholesky vectors, yielding the much less memory-intensive three-index Cholesky intermediates.~\citep{Schnack2022efficient} 
\vspace*{0.2cm}\\
These advantages generalize straightforwardly to QED-CCSD. Here, the derivative of the dressed two-electron integral matrix ${g}_{pqrs}$ can be expanded as
\begin{equation}
\label{eq:g-tilde-derivative}
    {g}_{pqrs}^{[1]} = g_{pqrs}^{e[1]} + d_{pq}^{[1]}d_{rs} + d_{pq}d_{rs}^{[1]} . 
\end{equation}
In this expression, the derivative of the dipole moment contains both electronic and nuclear contributions, and we can further expand this term as
\begin{equation}
\label{eq:dipole-derivative}
    d_{pq}^{[1]} = d_{pq}^{e[1]} + \frac{S_{pq}^{[1]}}{N_e}d_N + \frac{S_{pq}}{N_e}d_N^{(1)}.
\end{equation}
In the evaluation of the gradient, the derivatives $(\alpha\beta|J)^{(1)}$ and $(J|K)^{(1)}$ are calculated in terms of AO-shells. In particular, a given \textit{J} represents a pair of AOs $\gamma\delta$ and the integral program calculates the integrals as $(\text{AB}|\text{CD})$, where the $\alpha$ is in the shell A, the $\beta$ is in the B shell, and so on. Now, for the undressed integral $g^e_{\text{ABCD}}=(\text{AB}\vert\text{CD})^e$, the derivative is non-zero only when differentiating with respect to one of the atoms in the shell quartet. As a consequence, its derivative only has 12 non-zero components. This reduced dimensionality is exploited to reduce computational costs.\cite{Schnack2022efficient} However, the generalization of the algorithm to the QED case requires the introduction of the derivative of the dipole moment in eq.~\eqref{eq:dipole-derivative}. Here, we note that the first two terms have 6 non-zero components, while the third one, involving the derivative of the nuclear dipole moment, has 3N non-zero components, where N is the number of atoms in the molecule. As a consequence, the generalization to the QED case requires a redefinition of $(\alpha\beta|J)^{(1)}$ and $(K|L)^{(1)}$ to include the QED contributions with 6 non-zero components and a subsequent separate treatment of the $d_N^{(1)}$ contributions. This will also be the case for the two-electron part of the orbital relaxation gradient. 
\vspace*{0.2cm}\\
From eqs.~\eqref{eq:g-tilde-derivative} and \eqref{eq:dipole-derivative}, the $d_N^{(1)}$ contributions to the two-electron gradient are
\begin{equation}
\label{eq:nuclear-dipole-derivative}
    \frac{d_N^{(1)}}{N_e}(S_{pq}d_{rs}+d_{pq}S_{rs}) .
\end{equation}
In order to treat the $d_N^{(1)}$ contributions, we could insert this expression into eq.~\eqref{eq:2e-contraction-Cholesky-CCSD}.
However, this is equivalent to explicitly considering these terms without introducing the Cholesky decomposition. Here, we present only the final expression of the $d_N$ contributions to the two-electron gradient, while we show this equivalence in the Supporting Information.\\
From eq.~\eqref{eq:nuclear-dipole-derivative}, the additional terms needed for the two-electron gradient are:
\begin{equation}
    \frac12\frac{d_N^{(1)}}{N_e} \sum_{pqrs}d^e_{pqrs}  (S_{pq}d_{rs}+d_{pq}S_{rs}) = \frac{d_N^{(1)}}{N_e} \sum_p W_{pp}^{\text{dipole}}
    \label{eq:W_pp_dipole}
\end{equation}
where we have defined
\begin{equation}
    W_{pq}^{\text{dipole}} = \sum_{rs}d^e_{pqrs}d_{rs} . 
\end{equation}
Note that, by using
\begin{equation}
    \sum_p{d^e_{pprs}}  =  (N_e - 1) D^e_{rs},
\end{equation}
we can simplify eq.~\eqref{eq:W_pp_dipole} and rewrite it as
\begin{equation}
    \frac{d_N^{(1)}}{N_e} \sum_p W_{pp}^{\text{dipole}} = d_N^{(1)}\left(1-\frac{1}{N_{e}}\right)\sum_{pq}D^e_{pq} d_{pq}.
\end{equation}
Finally, using eq.~\eqref{eq:nuclear-dipole-derivative}, the nuclear contributions to the orbital relaxation terms read
\begin{equation}
\begin{split}
    \sum_{ai}\sum_j\Bar{\kappa}_{ai}(2g_{aijj}^{[1],N}-g_{ajji}^{[1],N}) &= \frac{d_N^{(1)}}{N_e}\sum_{aij} \bar{\kappa}_{ai} (2d_{ai}\delta_{jj} - d_{aj}\delta_{ji}) \\
    &=d_N^{(1)}\bigg(1-\frac{1}{N_{e}}\bigg)\sum_{ai}\Bar{\kappa}_{ai}d_{ai}.
\end{split}
\end{equation}
\subsubsection{Reorthonormalization}
\noindent
Above, we have focused on the UMO contributions to the molecular gradient. The last terms to account for are the reorthonomalization contributions, which arise from the derivative of the orbital connection. 
These contributions are usually expressed in the MO basis in terms of a generalized Fock matrix $\mathcal{F}_{pq}$ such that the gradient contribution takes the form $-\sum_{pq}S_{pq}^{[1]}\mathcal{F}_{pq}$. In the QED-CCSD-1 case, $\mathcal{F}_{pq}$ can be defined as
\begin{equation}
\label{eq:generalized-fock-matrix}
 \mathcal{F}_{pq}= \frac12(\mathscr{F}_{pq} + \mathscr{F}^{\langle d\rangle}_{p q} + \mathscr{F}^{\bar{\kappa},2e}_{pq}),
\end{equation}
where
\begin{equation}
\begin{split}
    &\mathscr{F}_{pq} =\sum_r\bigg[{\mathcal{D}^{\bar{\kappa}}_{p r} h_{r q}} +\sum_J \tilde{\mathcal{W}}^J_{pr}L_{qr}^J\bigg] +\frac12 \sum_{r s}{\mathcal{D}^{\bar{\kappa}}_{r s} d_{p r} d_{s q}} -\frac{ \langle d \rangle^2}{2 N_e}\mathcal{D}^{\bar{\kappa}}_{p q}+\sqrt{\frac{\omega}{2}} \sum_{r}{\mathcal{D}^{e\text{-}p}_{p r} d_{r q}} \\
    &\mathscr{F}^{\langle d\rangle}_{p i} =-4d_{p i}\bigg(\sum_{rs}{D^{\bar{\kappa}}_{rs} d_{rs}} - \langle d \rangle - \sqrt{\frac{\omega}{2}} D^p\bigg) \\
    &\mathscr{F}^{\langle d\rangle}_{p a}  = 0  \\
    &D_{pq}^{\bar{\kappa}} = D_{pq} + \bar{\kappa}_{pq} \hspace{100pt} \mathcal{D}_{pq}^X = D_{pq}^X + D_{qp}^X
\end{split}
\end{equation}
Here, we have separated the two-electron contributions to the orbital relaxation in $\mathscr{F}^{\bar{\kappa},2e}_{pq}$. Since the $d_N^{(1)}$ part of the derivative dipole moment has no reorthornormalization contributions, $\mathscr{F}^{\bar{\kappa},2e}_{pq}$ is the same as the one in the CCSD case, with dressed two-electron integrals. We refer to Ref.\citenum{Schnack2022efficient} for this term. The other two terms are given in the Supporting Information.
\section{4. Results and discussion}
To illustrate the efficiency and possible applications of the QED-CCSD-1 gradients, we present timings of the implementation as well as optimized geometries for a few molecular systems. 
The QED-CCSD-1 gradient has been implemented in a development branch of the eT 2.0 program.\cite{eT} Geometry optimizations were performed using an interface to the geomeTRIC package,\cite{wang2016geometry} which allows for efficient optimization of molecular geometries, as well as global orientation, which is crucial when optimizing molecular systems in the presence of an external field.\cite{liebenthal2024orientation} For the QED-CCSD-1 calculations, the cavity frequency is set equal to the first bright excitation energy, calculated at the CCSD level. The coupling strength is set to $\vert\vert\boldsymbol{\lambda}\vert\vert = 0.05$ a.u. 
All geometry optimizations were performed using the cc-pVDZ basis set. The coordinates for the optimized geometries are provided in a separate repository.\cite{zenodo} 
Timings were run on an Intel(R) Xeon(R) Platinum 8380 system with 80  cores and 2 TB of memory.
\subsection{4A. Timings}
We compare timings for the evaluation of the gradient with CCSD and QED-CCSD-1 for three molecular systems. 
As test systems, we consider cyclooctatetraene, with the aug-cc-pVTZ basis set, and azobenzene and the porphine molecule with the cc-pVDZ basis sets. The selected molecules are shown in Figures \ref{fig:cyclo}, \ref{fig:azobenzene_cis}, and \ref{fig:porphine}. Timing data are presented in Table \ref{tab:timing}. The time to solve the multiplier equations [eq.~\eqref{eq:multipliers}] is not included in the gradient time as this is not strictly part of the gradient evaluation, but it represents a significant part of the total time of the calculation.
\vspace*{0.2cm}\\
As expected, we see that in all cases, the time for the molecular gradient evaluation is longer for QED-CCSD-1 than for CCSD. However, the evaluation of the gradient always constitutes a very small fraction of the total time. For example, in cyclooctatetraene, the evaluation of the analytical molecular gradient only represents 8\% of the time of the whole calculation. 
In the evaluation of the QED-CCSD-1 gradient, the additional cost compared to CCSD is found to be almost exclusively due to the QED terms in the two-electron density matrix. The remaining QED contributions add a negligible cost compared to CCSD.
\begin{table}[!htbp]
    \centering
    \caption{Timings for a single gradient evaluation using CCSD and QED-CCSD-1. }
    \begin{tabular}{l l l c c c c}
        \toprule
         & \hspace{0.2cm}$n_\text{occ}$/$n_\text{vir}$ & Method & $t_{\text{2e-dens}}$ & $t_{\text{gradient}}$ & $t_{\text{total}}$ & $t_{\text{gradient}}/t_{\text{total}}$  \\
        \midrule
        Cyclooctatetraene   & \multirow{2}{2cm}{\hspace{0.2cm} 28/524}  & CCSD      & 243 s  & 324 s  & 3154 s  & (10.3 \%)  \\
        (aug-cc-pVTZ)       &                                           & QED-CCSD-1  & 522 s  & 611 s  & 7959 s  & (7.9 \%)   \\ \\
        Azobenzene          & \multirow{2}{2cm}{\hspace{0.2cm} 48/198}  & CCSD      & 9.7 s  & 22.9 s & 267.1 s & (8.6 \%)   \\
        (cc-pVDZ)           &                                           & QED-CCSD-1  & 25.7 s & 41.9 s & 872.1 s & (4.8 \%)   \\ \\
        Porphine            & \multirow{2}{2cm}{\hspace{0.2cm} 81/325}  & CCSD      & 140 s  & 267 s  & 3303 s  & (8.1 \%)   \\
        (cc-pVDZ)           &                                           & QED-CCSD-1  & 406 s  & 530 s  & 13821 s & (3.8 \%)   \\
        \bottomrule
    \end{tabular}
    \label{tab:timing}
\end{table}
\subsection{4B. Geometry optimization}
\subsubsection{Cyclooctatetraene}
\begin{figure}[!htbp]
    \includegraphics[width=0.45\textwidth]{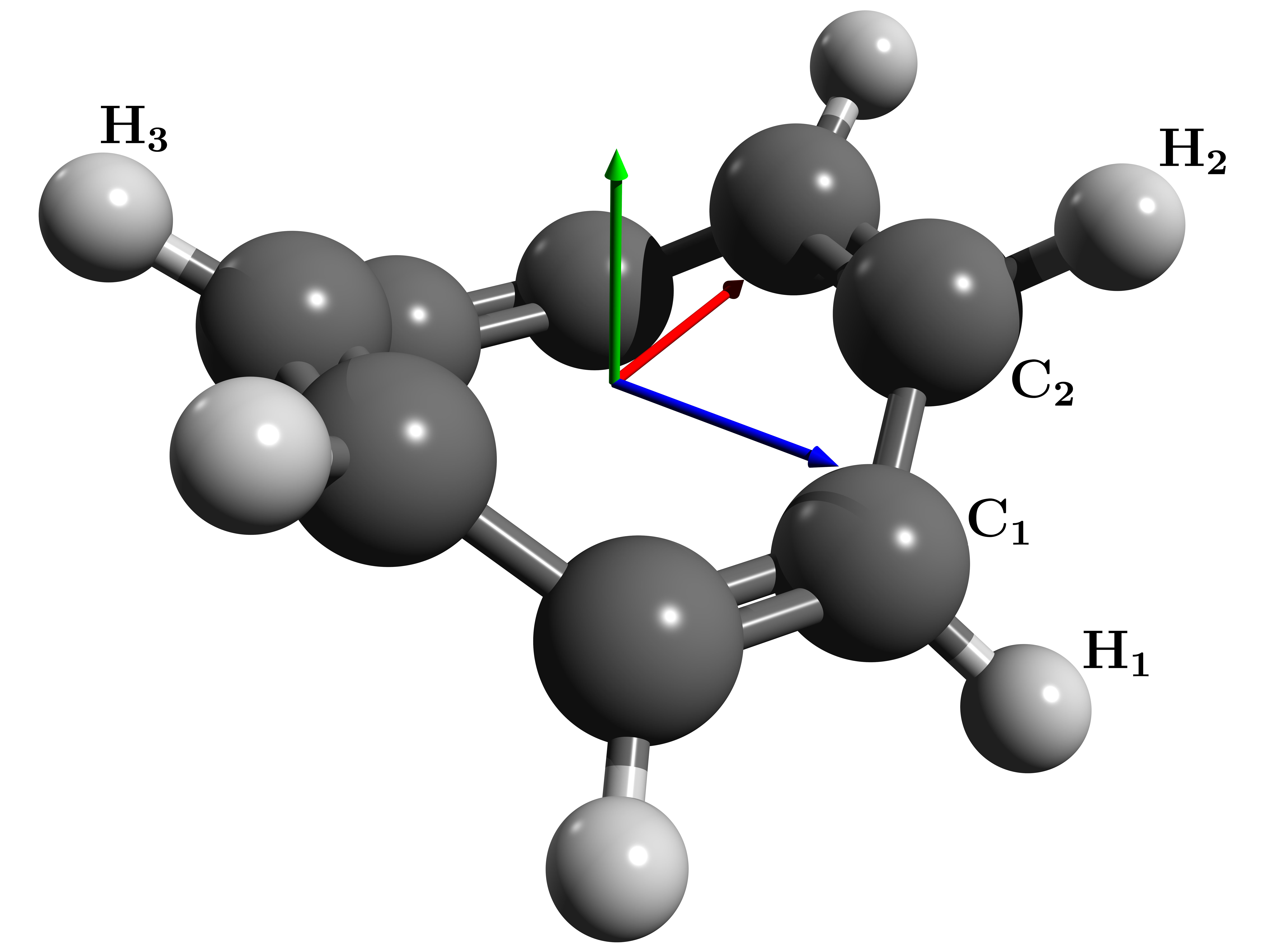}
    \caption{Optimized geometry for the cyclooctatetraene molecule using QED-CCSD-1 with $\lambda = 0.05$ a.u. and the cc-pVDZ basis set. The polarization vector is indicated by the green arrow.}
    \label{fig:cyclo}
\end{figure}

The anti-aromatic molecule cyclooctatetraene has a boat shape in its ground state geometry. 
Starting from the optimized CCSD geometry in a random orientation, we find that the field causes the molecule to reorient such that the plane of the boat lies perpendicular to the cavity polarization. This allows the molecule to minimize its spatial extent along the polarization axis, which lowers the total energy. The optimized geometry is shown in Figure \ref{fig:cyclo}. As expected, in addition to the reorientation, we see a slight flattening of the molecule along the polarization axis. 
In particular, we find a slight increase in the distance between two opposite hydrogen atoms (H$_2$ and H$_3$) and a slight decrease in the dihedral angle $\gamma$ between two adjacent CH groups (H$_1$, C$_1$, C$_2$, H$_2$). Values of these selected bond lengths and angles are given in Table~\ref{tab:cyclo}. 
\begin{table}[!htbp]
    \centering
    \caption{Selection of optimized parameters for cyclooctatetraene. $\gamma$ is the dihedral angle between H$_1$-C$_1$-C$_2$-H$_2$.}
    \begin{tabular}{l C{3cm} C{3cm}}
        \toprule
                Method        & r$_{H_2-H_3}$ [Å] & $\gamma$ [$^\circ$]   \\
        \midrule
           CCSD               & 5.22  &   48.4       \\
         QED-CCSD-1           & 5.25  & 47.1      \\
        \bottomrule
    \end{tabular}
    \label{tab:cyclo}
\end{table}
\subsubsection{Azobenzene}
\begin{figure}[!htbp]
    \includegraphics[width=0.5\textwidth]{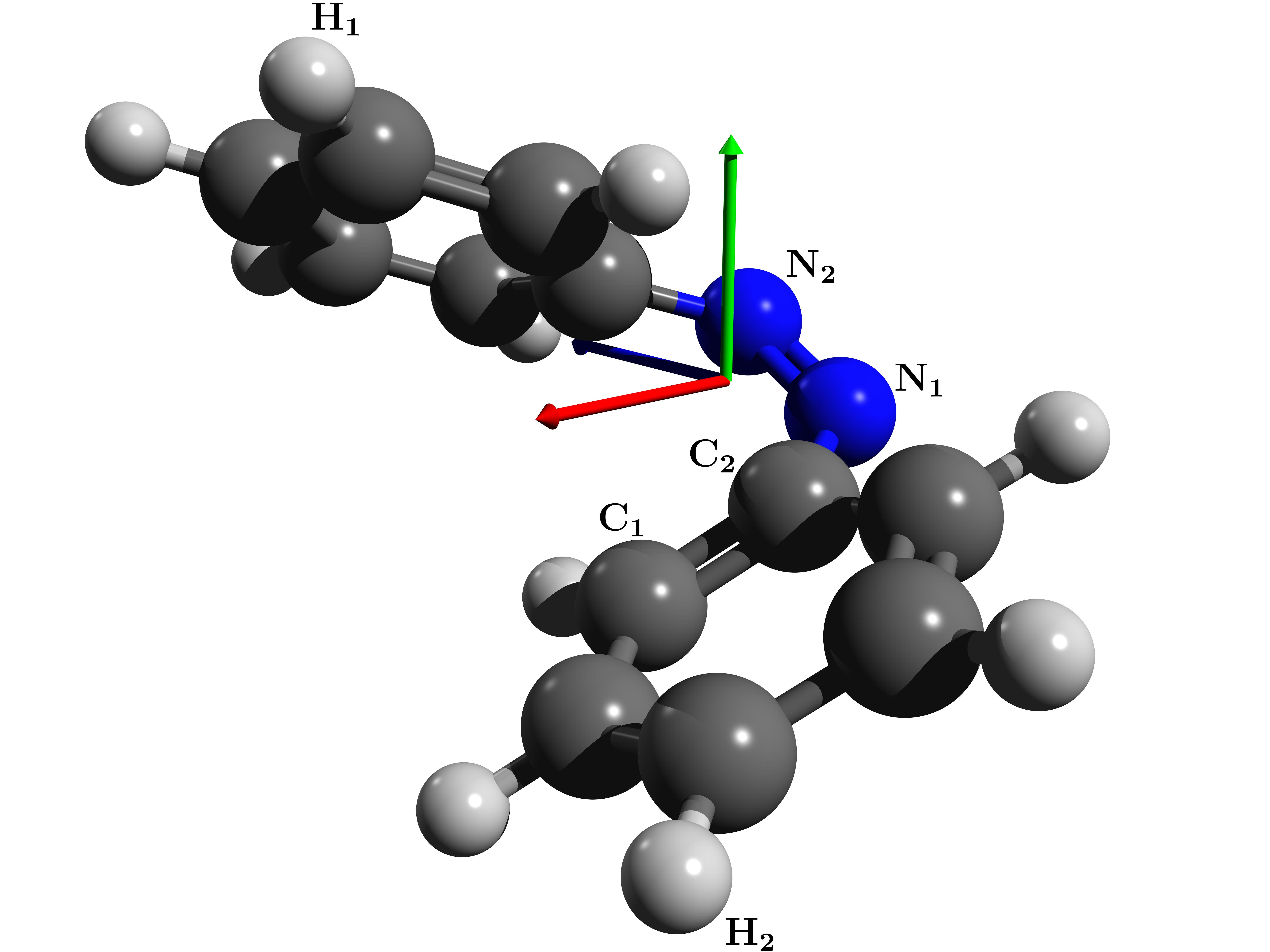}
    \caption{Optimized geometry for the cis-isomer of azobenzene using QED-CCSD-1 with $\lambda = 0.05$ a.u. and the cc-pVDZ basis set. The polarization vector is indicated by the green arrow.}
    \label{fig:azobenzene_cis}
\end{figure}

As a second example, we consider the cis-isomer of azobenzene.
As in the previous case, starting from the optimized CCSD geometry in a random orientation, we find that the molecule rotates in order to align to the cavity polarization. As can be seen in Figure \ref{fig:azobenzene_cis}, however, the optimized molecular structure preserves a certain angle relative to the polarization direction. In Table~\ref{tab:azocis}, we report this relative orientation of the molecule as the angle $\theta$ between the N-N bond and the direction of the cavity field.
When introducing the cavity field, we also observe a rotation of the phenyl groups around the C-N bonds, which leads the groups to be more perpendicular to the polarization. In this case, we report in Table~\ref{tab:azocis} both the angle between the plane of the phenyl group and the polarization direction $\phi$ and the dihedral angle $\alpha$ between  C$_1$, C$_2$, N$_1$ and N$_2$ (see Figure \ref{fig:azobenzene_cis}).
\begin{table}[!htbp]
    \centering
    \caption{Selection of optimized internal coordinates for cis-azobenzene. $\alpha$ is the C$_1$-C$_2$-N$_1$-N$_2$ dihedral angle, $\theta$ is the angle between the N-N bond and the polarization vector, and $\phi$ is the angle between the C$_1$-C$_2$ phenyl group and the polarization direction.}
    \begin{tabular}{l C{3cm} C{3cm} C{3cm} C{3cm}}
        \toprule
          Method        & r$_{H_1-H_2}$ [Å]  & $\alpha$ [$^\circ$] &  $\theta$ [$^\circ$] &   $\phi$ [$^\circ$]\\ 
        \midrule
         CCSD             & 5.60  &57.1 & - & -  \\ 
         QED-CCSD-1      & 5.19  & 52.0 & 61.3  &30.7  \\
        \bottomrule
    \end{tabular}
    \label{tab:azocis}
\end{table}
\subsubsection{Porphine}
\begin{figure}[!htbp]
    \includegraphics[width=0.5\textwidth]{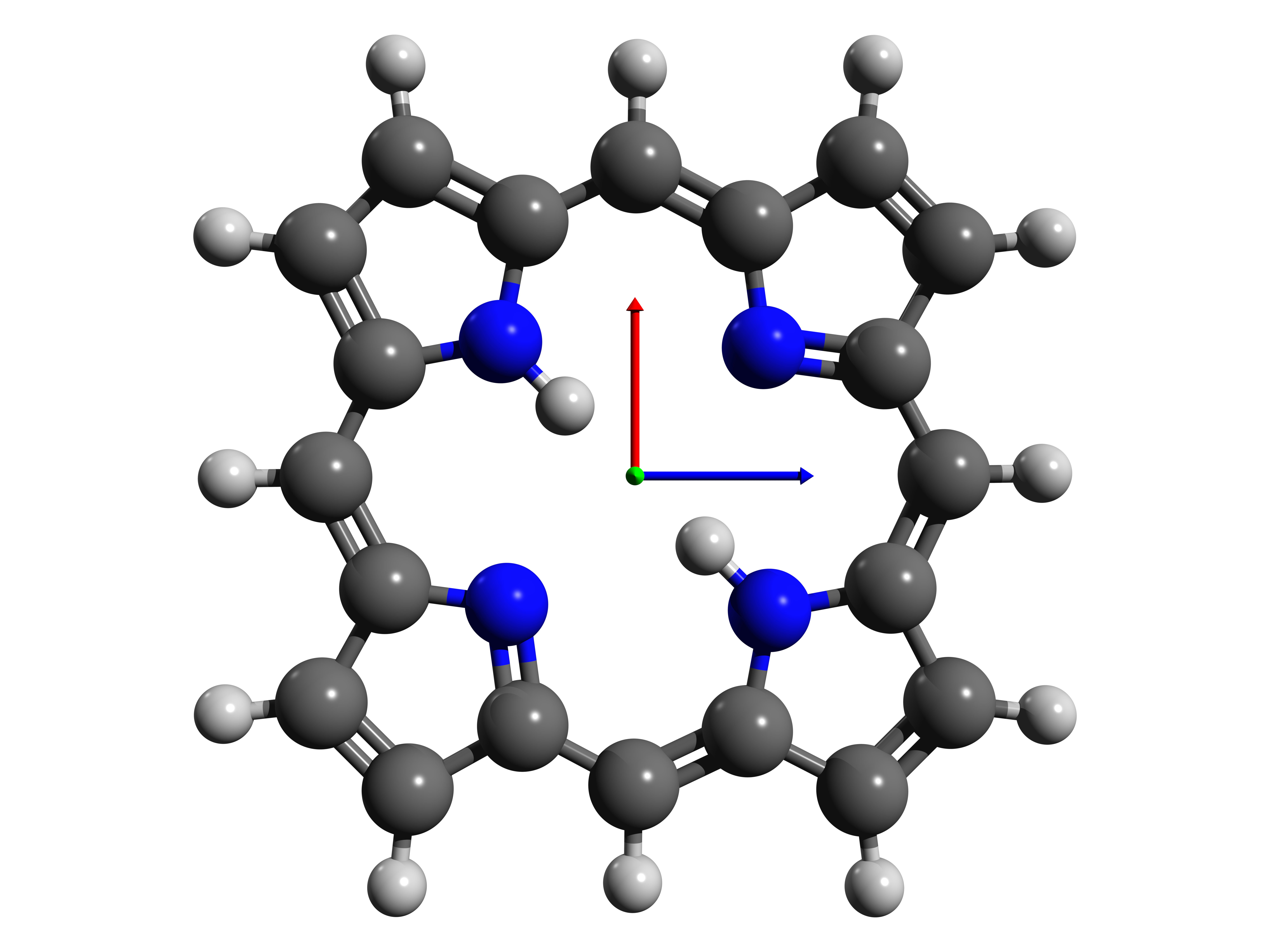}
    \caption{Optimized geometry for the porphine molecule using QED-CCSD-1 with $\lambda = 0.05$ a.u. and the cc-pVDZ basis set. The molecule is planar, with the polarization vector perpendicular to the plane of the molecule, indicated by the green arrow pointing out of the figure.}
    \label{fig:porphine}
\end{figure}

Porphine is the base-structure for large categories of biological molecules called porphyrins and chlorins.
The optimized geometry is shown in Figure \ref{fig:porphine}. Again, we find that the molecule orients itself to become perpendicular to the polarization vector. In contrast to the other two systems, however, the internal geometry of this molecule does not change noticeably when coupling to the field. A different behavior, however, might be obtained considering more than one cavity mode with different polarization directions.
\section{5. Conclusions}
In this paper, we presented a formulation of ground state analytical gradients for QED-CC together with an efficient Cholesky-based implementation at the QED-CCSD-1 level.
Using the resolution-of-identity form, we avoid the evaluation of the derivative of the Cholesky vectors.\cite{delcey2014analytical,Schnack2022efficient} Moreover, building on an existing implementation of CCSD gradients,\cite{Schnack2022efficient} we reduce the memory usage by using an on-the-fly construction of the intermediates involving the V$^4$ block of the density matrix. 
Timings for a single gradient evaluation show that the calculation of the analytical gradient requires less than 10\% of the total time of the QED-CCSD-1 calculation. 
\vspace*{0.2cm}\\
Moreover, we optimized the geometries of cyclooctatetraene, azobenzene, and porphine in an optical cavity. In all cases, we allowed for rotations of the molecules, thus showing the reorientation of the system with respect to the polarization of the field. 
This highlights the importance of including cavity-induced effects when determining optimal geometries, as already suggested by some recent studies.\cite{castagnola2024polaritonic,liebenthal2024orientation,schnappinger2024molecular} 
\vspace*{0.2cm}\\
Given the well-established accuracy of coupled cluster theory and the efficiency of the implementation, we believe that the implementation will prove to be a useful tool for determining equilibrium geometries in optical cavities. Moreover, the implementation can also be used to parametrize classical force fields or to perform ground state \emph{ab initio} molecular dynamics simulations of cavity-induced orientational effects in molecular ensembles.
\section*{Acknowledgements}
We thank Tor S. Haugland and Matteo Castagnola for insightful discussions and Matteo Castagnola for useful comments on the draft. This work was supported by the European Research Council (ERC) under the European Union's Horizon 2020 Research and Innovation Programme (grant agreement No. 101020016). 
\section*{Supporting Information}
Comparison with numerical gradients, notes on the treatment of the derivative of the nuclear dipole moments, derivations of the reorthonormalization terms and programmable expressions for the densities.
\bibliography{bibliography}


\section*{Supplementary material}

**Supporting information for "Analytical evaluation of ground state gradients in quantum electrodynamics coupled cluster theory"**

Marcus T. Lexander,[a] Sara Angelico,[a] Eirik F. Kjønstad, and Henrik Koch

*Department of Chemistry, Norwegian University of Science and Technology, 7491 Trondheim*

(Dated: 12 June 2024)

---

[a] These authors contributed equally to this work

# CONTENTS

# S1. COMPARISON OF THE ANALYTICAL AND NUMERICAL GRADIENTS

To test the correctness of the implementation of the analytical gradient, we compare the analytical and the numerical gradients for $H_2O$-He. The geometry of the system can be found in Table S1. The gradient was calculated with the cc-pVDZ basis set. The frequency, coupling strength and polarization directions are $\omega = 0.5$ a.u., $\lambda = 0.05$ a.u., $\epsilon = [0.577350, 0.577350, 0.577350]$. The numerical gradient is calculated with a five point stencil, with $h = 1 \cdot 10^{-4}$ Å. The maximum deviation between the analytical and numerical gradient is $4 \cdot 10^{-9}$.

TABLE S1. Comparison of numerical and analytical gradients. The gradients are calculated using the cc-pVDZ basis set. Numerical gradients were evaluated with a five-point stencil using a displacement of $h = 1 \cdot 10^{-4}$ Å. The molecular geometry is given in Å, while the gradients are expressed in atomic units. The a.u. to Å conversion factor used is 0.52917721092.

|  | Molecular geometry (Å) | | |
| --- | --- | --- | --- |
| H  | 0.866 810 000   | 0.601 440 000  | 5.000 000 000 |
| H  | −0.866 810 000  | 0.601 440 000  | 5.000 000 000 |
| O  | 0.000 000 000   | −0.075 790 000 | 5.000 000 000 |
| He | 0.100 000 000   | −0.020 000 000 | 7.530 000 000 |
|  | Analytical gradient (a.u.) | | |
| H  | 0.074 446 196   | 0.047 170 451  | 0.000 360 463 |
| H  | −0.074 310 167  | 0.046 932 779  | 0.000 053 345 |
| O  | −0.000 031 410  | −0.094 104 796 | 0.002 633 639 |
| He | −0.000 104 618  | 0.000 001 566  | −0.003 047 447 |
|  | Numerical gradient (a.u.) | | |
| H  | 0.074 446 196   | 0.047 170 450  | 0.000 360 459 |
| H  | −0.074 310 166  | 0.046 932 781  | 0.000 053 346 |
| O  | −0.000 031 410  | −0.094 104 796 | 0.002 633 640 |
| He | −0.000 104 614  | 0.000 001 565  | −0.003 047 451 |

## S2. NOTES ON THE TREATMENT OF THE NUCLEAR CONTRIBUTIONS TO THE CHOLESKY VECTORS

In the following, we will make use of the following identities:

$$\sum_{JK} S_J(J|K)^{-1}(K|pq) = S_{pq} = \delta_{pq} \qquad \sum_{JK} d_J(J|K)^{-1}(K|pq) = d_{pq} \qquad (1)$$

We will now prove the first identity. A similar derivation can be followed to obtained the second one.

To prove the first identity, we start by considering $W_K = \sum_J S_J(J|K)^{-1}$. Inverting this relation, we get

$$\sum_K W_K(K|L) = S_L. \qquad (2)$$

Using that $L$ corresponds to a pair of AOs $\gamma\delta$ and transforming them to the MO basis we obtain

$$\sum_K \sum_{\gamma\delta} W_K S_{K,\gamma\delta} C_{\gamma p} C_{\delta q} = S_{pq} \qquad (3)$$

$$\sum_K W_K S_{K,pq} = \delta_{pq}. \qquad (4)$$

Substituting the definition of $W_K$ we find

$$\sum_{KJ} S_J(J|K)^{-1}(K|pq) = \delta_{pq} \qquad (5)$$

We can now consider the $d_N^{(1)}$ contributions to the two-electron gradient. As shown in eq. (48), these can be expressed as

$$\frac{1}{2}\sum_{pqrs} d^e_{pqrs} g^{[1]}_{pqrs} = \sum_{pqJ} (pq|J)^{[1]} \left( \sum_{rsK} d_{pqrs}(J|K)^{-1}(K|rs) \right)$$
$$- \frac{1}{2} \sum_{JK} \left( \sum_{pqrs} \sum_{ML} d_{pqrs}(pq|J)(J|M)^{-1}(M|L)^{(1)}(L|K)^{-1}(K|rs) \right). \qquad (6)$$

We can now notice that the $d_N^{(1)}$ contributions to $(pq|J)^{[1]}$ and $(M|L)^{(1)}$ are:

$$(pq|J)^{[1]} = \frac{d_N^{(1)}}{N_e}(S_{pq}d_J + d_{pq}S_J) \qquad (7)$$

$$(M|L)^{(1)} = \frac{d_N^{(1)}}{N_e}(S_M d_L + d_M S_L). \qquad (8)$$

Then, inserting eq. (7) into the first term of eq. (6), one gets

$$\frac{d_N^{(1)}}{N_e} \sum_{pqrs} d_{pqrs} \sum_{JK} (S_{pq}d_J + d_{pq}S_J)(J|K)^{-1}(K|rs). \tag{9}$$

Now, using the identities in eq. (1), we find

$$\frac{d_N^{(1)}}{N_e} \sum_{pqrs} d_{pqrs}(\delta_{pq}d_{rs} + d_{pq}\delta_{rs}) = 2\frac{d_N^{(1)}}{N_e} \sum_p W_{pp}^{\text{dipole}}, \tag{10}$$

where we have used the symmetry properties of $d_{pqrs}$ and defined $W_{pq}^{\text{dipole}} = \sum_{rs} d_{pqrs}d_{rs}$.

Inserting eq. (8) in the second term of eq. (6), one finds

$$-\frac{1}{2}\frac{d_N^{(1)}}{N_e} \sum_{pqrs} d_{pqrs} \sum_{MJ}\sum_{LK} (pq|J)(J|M)^{-1}(S_M d_L + d_M S_L)(L|K)^{-1}(K|rs). \tag{11}$$

Applying the identities in eq. (1) one finds

$$-\frac{1}{2}\frac{d_N^{(1)}}{N_e} \sum_{pqrs} d_{pqrs}(\delta_{pq}d_{rs} + d_{pq}\delta_{rs}) = -\frac{d_N^{(1)}}{N_e} \sum_p W_{pp}^{\text{dipole}}. \tag{12}$$

Finally, the sum of the two terms gives $\frac{d_N^{(1)}}{N_e} \sum_p W_{pp}^{\text{dipole}}$. Note that this term is equivalent to the one derived without considering the Cholesky decomposition of the dressed two-electron integrals proposed in the same text.

As for the two-electron integrals contributions to the orbital relaxation terms, these can be written as:[1]

$$\sum_{aij} \bar{\kappa}_{ai}(2g^{[1]}_{aijj} - g^{[1]}_{ajji}) = \sum_{aiK} K_{ai}^K (ai|K)^{[1]} + \sum_{jK} L^K (K|jj)^{[1]} + \sum_{jiK} N_{ji}^K (K|ji)^{[1]} \tag{13}$$

$$-\sum_{KL} M_{KL}(K|L)^{(1)} - \sum_{KL} O_{KL}(K|L)^{(1)} \tag{14}$$

where we have used the intermediates[1]

$$K_{ai}^K = \sum_j (2\bar{\kappa}_{ai}Z_{jj}^K - \bar{\kappa}_{aj}Z_{ij}^K) \quad N_{ji}^K = -\sum_a \bar{\kappa}_{ai}Z_{aj}^K$$

$$L^K = 2\sum_{ai} \bar{\kappa}_{ai}Z_{ai}^K$$

$$M_{KL} = \sum_j L^K Z_{jj}^L \qquad O_{KL} = \sum_{ij} N_{ji}^K Z_{ji}^L.$$

Using the definitions of the different intermediates and only considering the $d_N^{(1)}$ contributions to the derivatives, we get for the first term:

$$\frac{d_N^{(1)}}{N_e} \left( \sum_{aij} 2\bar{\kappa}_{ai}d_{ai} \sum_K S_K Z_{jj}^K - \sum_{aij} \bar{\kappa}_{aj}d_{ai} \sum_K S_K Z_{ij}^K \right) \tag{15}$$

$$= \frac{d_N^{(1)}}{N_e} \left( N_e \sum_{ai} \bar{\kappa}_{ai}d_{ai} - \sum_{ai} \bar{\kappa}_{ai}d_{ai} \right) = d_N^{(1)} \left(1 - \frac{1}{N_e}\right)\left(\sum_{ai} \bar{\kappa}_{ai}d_{ai}\right) \tag{16}$$

while for the second term:

$$2\frac{d_N^{(1)}}{N_e}\left(\sum_{aij}\bar{\kappa}_{ai}d_{jj}\sum_K S_K Z_{ai}^K + \sum_{aij}\bar{\kappa}_{ai}\delta_{jj}\sum_K d_K Z_{ai}^K\right) = d_N^{(1)}\sum_{ai}\bar{\kappa}_{ai}d_{ai}. \quad (17)$$

For the third term:

$$-\frac{d_N^{(1)}}{N_e}\left(\sum_{aij}\bar{\kappa}_{ai}d_{ji}\sum_K S_K Z_{aj}^K + \sum_{aij}\bar{\kappa}_{ai}S_{ji}\sum_K d_K Z_{aj}^K\right) = -\frac{d_N^{(1)}}{N_e}\sum_{ai}\bar{\kappa}_{ai}d_{ai}. \quad (18)$$

For the fourth term:

$$-2\frac{d_N^{(1)}}{N_e}\sum_{aij}\bar{\kappa}_{ai}\left(\sum_{KL} Z_{ai}^K S_K d_L Z_{jj}^L + \sum_{KL} Z_{ai}^K d_K S_L Z_{jj}^L\right) = -d_N^{(1)}\sum_{ai}\bar{\kappa}_{ai}d_{ai}. \quad (19)$$

Finally, for the fifth term we find:

$$\frac{d_N^{(1)}}{N_e}\sum_{aij}\bar{\kappa}_{ai}\left(\sum_{KL} Z_{aj}^K S_K d_L Z_{ji}^L + \sum_{KL} Z_{aj}^K d_K S_L Z_{ji}^L\right) = \frac{d_N^{(1)}}{N_e}\sum_{ai}\bar{\kappa}_{ai}d_{ai}. \quad (20)$$

Where we have made extensive use of the identities in eq. (1). Note that the second and the fourth terms cancels each other out, as well as the third and the fifth. As a consequence, the $d_N^{(1)}$ contributions to the orbital relaxation gradient can be expressed as

$$\sum_{aij}\bar{\kappa}_{ai}(2g_{aijj}^{[1]} - g_{ajji}^{[1]}) = d_N^{(1)}\left(1 - \frac{1}{N_e}\right)\left(\sum_{ai}\bar{\kappa}_{ai}d_{ai}\right) \quad (21)$$

which is equivalent (up to the chosen threshold for the Cholesky decomposition) to the expression obtained without differentiating the Cholesky decomposition.

## S3. DERIVATIONS FOR THE REORTHONORMALIZATION CONTRIBUTIONS

The reorthonormalization terms in the gradient are defined as the contributions arising from the one-index transformations of the Hamiltonian and the Fock matrix. This is defined as:

$$h_{pq}^{\{1\}} = \{S^{[1]}, h\}_{pq} = \sum_t S_{pt}^{[1]} h_{tq} + S_{qt}^{[1]} h_{pt}. \quad (22)$$

In the evaluation of the gradient, terms of this kind are contracted with the densities or the orbital relaxation multipliers $\bar{\kappa}_{ai}$, and we can rewrite these contributions as expressions of the form $-\sum_{pq} S_{pq}^{[1]}\mathcal{F}_{pq}$. In the following, we provide derivations of the first two terms of the generalized Fock matrix presented in the main text in eq. (59). Expressions of $\mathcal{F}_{pq}^{\bar{\kappa},2e}$ can be found in Ref. 1 and are omitted here.

## A. Dressed one-electron integral

The dressed one-electron integral is defined as

$$h_{pq} = h_{pq}^e + \frac{1}{2}\sum_r d_{pr}d_{rq} - \langle d\rangle d_{pq} + \frac{\delta_{pq}}{2N_e}\langle d\rangle^2. \qquad (23)$$

In the following, we will analyze all the contributions coming from the different terms of this integral. We will use the notation $\mathscr{F}_{pq}\mathrel{+}=$ to indicate contributions to add to $\mathscr{F}_{pq}$. The reorthonormalization contributions for the purely electronic part are derived as:

$$\sum_{pq} D_{pq}^e h_{pq}^{\{1\}} = \sum_{pqt}\left(D_{pq}^e S_{pt}^{[1]} h_{tq} + D_{pq}^e S_{qt}^{[1]} h_{pt}\right) = \sum_{pq} S_{pq}^{[1]}\left(\sum_r \mathcal{D}_{pr}^e h_{rq}\right) \qquad (24)$$

$$\Rightarrow \mathscr{F}_{pq}\mathrel{+}= \left(\sum_r \mathcal{D}_{pr}^e h_{rq}\right)$$

where in the last equality we have redefined $p\leftrightarrow q$ and used the symmetry of $h_{pq}$. Moreover, we have introduced the symmetrized density matrix $\mathcal{D}_{pq}^e = D_{pq}^e + D_{qp}^e$.

For the term coming from the dipole self-energy $\frac{1}{2}\sum_r d_{pr}d_{rq}$ we have:

$$\left(\sum_r d_{pr}d_{rq}\right)^{\{1\}} = \sum_r\left(d_{pr}d_{rq}^{\{1\}} + d_{pr}^{\{1\}}d_{rq}\right) \qquad (25)$$

$$\sum_{pq} D_{pq}^e\left(\sum_r d_{pr}d_{rq}\right)^{\{1\}} = \sum_{pqr}\left(D_{pq}^e d_{pr}d_{rq}^{\{1\}} + D_{pq}^e d_{pr}^{\{1\}}d_{rq}\right) = \sum_{pqr}\mathcal{D}_{pq}^e d_{pr}d_{rq}^{\{1\}}$$

$$= \sum_{pqrt}\mathcal{D}_{pq}^e d_{pr}S_{rt}^{[1]}d_{tq} + \mathcal{D}_{pq}^e d_{pr}S_{qt}^{[1]}d_{rt}$$

$$= \sum_{pq} S_{pq}^{[1]}\left(\sum_{rs}\mathcal{D}_{rs}^e d_{pr}d_{sq} + \mathcal{D}_{pr}^e d_{rs}d_{sq}\right) \qquad (26)$$

$$\Rightarrow \mathscr{F}_{pq}\mathrel{+}= \sum_{rs}\mathcal{D}_{rs}^e d_{pr}d_{sq} + \sum_{rs}\mathcal{D}_{pr}^e d_{rs}d_{sq}.$$

For the $\langle d\rangle d_{pq}$ term, instead, we get:

$$(\langle d\rangle d_{pq})^{\{1\}} = \langle d\rangle^{\{1\}} d_{pq} + \langle d\rangle d_{pq}^{\{1\}}$$

$$\sum_{pq} D_{pq}^e(\langle d\rangle d_{pq})^{\{1\}} = \sum_{pq} D_{pq}^e \langle d\rangle^{\{1\}} d_{pq} + \sum_{pq} D_{pq}^e \langle d\rangle d_{pq}^{\{1\}}. \qquad (27)$$

S7

From the first term we get

$$\langle d \rangle^{\{1\}} = \sum_i 2d_{ii}^{\{1\}} = 2\sum_i \{S^{[1]}, d\}_{ii} = 2\sum_{it} S_{it}^{[1]} d_{ti} + S_{it}^{[1]} d_{it} = 4\sum_{it} S_{ti}^{[1]} d_{ti}$$

$$\sum_{pq} D_{pq}^e \langle d \rangle^{\{1\}} d_{pq} = \sum_{it} S_{ti}^{[1]} \left( 4d_{ti} \sum_{pq} D_{pq}^e d_{pq} \right) \quad \Rightarrow \quad \mathscr{F}_{pi}^{\langle d \rangle} \mathrel{+}= 4d_{pi} \sum_{rs} D_{rs}^e d_{rs}.$$

(28)

While from the second term we get

$$\sum_{pq} D_{pq}^e \langle d \rangle d_{pq}^{\{1\}} = \sum_{pq} S_{pq}^{[1]} \left( \langle d \rangle \sum_r \mathcal{D}_{pr}^e d_{rq} \right) \quad \Rightarrow \quad \mathscr{F}_{pq} \mathrel{+}= \langle d \rangle \sum_r \mathcal{D}_{pr}^e d_{rq}. \quad (29)$$

Finally, for the $\langle d \rangle^2$ term we have

$$\frac{\delta_{pq}}{2N_e}(\langle d \rangle^2)^{\{1\}} = \frac{\delta_{pq}}{N_e}\langle d \rangle \langle d \rangle^{\{1\}} = 4\frac{\delta_{pq}}{N_e}\langle d \rangle \sum_{it} S_{ti}^{[1]} d_{ti}$$

$$\sum_{pq} D_{pq}^e \frac{\delta_{pq}}{2N_e}(\langle d \rangle^2)^{\{1\}} = \sum_p \frac{D_{pp}^e}{2N_e}(\langle d \rangle^2)^{\{1\}} = \frac{1}{2}(\langle d \rangle^2)^{\{1\}} = \langle d \rangle \langle d \rangle^{\{1\}} = 4\langle d \rangle \sum_{it} S_{ti}^{[1]} d_{ti} \quad (30)$$

$$\Rightarrow \mathscr{F}_{pi}^{\langle d \rangle} \mathrel{+}= 4\langle d \rangle d_{pi}.$$

Collecting the different terms, we can now write

$$\mathscr{F}_{pq} = \sum_r \mathcal{D}_{pr}^e h_{rq} + \frac{1}{2}\sum_{rs} \mathcal{D}_{rs}^e d_{pr} d_{sq} - \frac{\mathcal{D}_{pq}^e}{2N_e}\langle d \rangle^2$$

$$\mathscr{F}_{pi}^{\langle d \rangle} = -4d_{pi}\sum_{rs} D_{rs}^e d_{rs} + 4\langle d \rangle d_{pi}$$

(31)

where

$$\sum_r \mathcal{D}_{pr}^e h_{rq} = \sum_r \mathcal{D}_{pr}^e h_{rq}^e + \frac{1}{2}\sum_{rs} \mathcal{D}_{pr}^e d_{rs} d_{sq} - \langle d \rangle \sum_r \mathcal{D}_{pr}^e d_{rq} + \frac{\mathcal{D}_{pq}^e}{2N_e}\langle d \rangle^2. \quad (32)$$

## B. Reorthonormalization of the bilinear term

The bilinear term in the Hamiltonian $\sqrt{\frac{\omega}{2}}\sum_{pq} d_{pq} E_{pq}(b^\dagger + b)$ leads to reorthonormalization terms that are:

$$\sqrt{\frac{\omega}{2}}\sum_{pq} D_{pq}^{e\text{-}p} d_{pq}^{\{1\}} \Rightarrow \mathscr{F}_{pq} \mathrel{+}= \sqrt{\frac{\omega}{2}}\sum_r \mathcal{D}_{pr}^{e\text{-}p} d_{rq} \quad (33)$$

and

$$-\sqrt{\frac{\omega}{2}} D^p \langle d \rangle^{\{1\}} = -4\sqrt{\frac{\omega}{2}} D^p \sum_{it} S_{ti}^{[1]} d_{ti} \Rightarrow \mathscr{F}_{pi}^{\langle d \rangle} \mathrel{+}= -4\sqrt{\frac{\omega}{2}} D^p d_{pi} \quad (34)$$

where we have reused the results for $d_{pq}^{\{1\}}$ and $\langle d \rangle^{\{1\}}$ derived in the previous section.

## C. Orbital relaxation (Fock matrix)

Finally, we derive the reorthonormalization contributions coming from the derivative of the Fock matrix. As the dressed two-electron integrals are decomposed in terms of Cholesky vectors, their contributions are automatically included using the implementation proposed in Ref. 1 with a proper redefinition of the Cholesky vectors. For this reason, we here focus only on the terms arising from the one-electron integrals.

Here, we define an extended matrix $\bar{\kappa}$ as:

$$\bar{\kappa} = \begin{pmatrix} 0 & 0 \\ \bar{\kappa}_{ai} & 0 \end{pmatrix} \tag{35}$$

We can now derive the reorthonormalization terms as:

$$\sum_{ai} \bar{\kappa}_{ai} h^{\{1\}}_{ai} = \sum_{pq} \bar{\kappa}_{pq} h^{\{1\}}_{pq} \tag{36}$$

And, applying the same procedure we used for the one-electron integrals contributions we get:

$$\begin{aligned} \mathscr{F}_{pq} &+= \sum_r \bar{\kappa}^s_{pr} h_{rq} + \frac{1}{2} \sum_{rs} \bar{\kappa}^s_{rs} d_{pr} d_{sq} - \frac{\bar{\kappa}^s_{pq}}{2N_e} \langle d \rangle^2 \\ \mathscr{F}^{\langle d \rangle}_{pi} &+= -4 d_{pi} \sum_{rs} \bar{\kappa}_{rs} d_{rs} + 4 \sum_p \frac{\bar{\kappa}_{pp}}{N_e} \langle d \rangle d_{pi} \end{aligned} \tag{37}$$

Here, we have introduced a symmetrized $\bar{\kappa}$ defined as $\bar{\kappa}^s_{pq} = \bar{\kappa}_{pq} + \bar{\kappa}_{qp}$, using the definition of $\bar{\kappa}$ given in eq. (35).

We can now note that $\bar{\kappa}_{pp} = 0$ and define some extended densities $D^{\bar{\kappa}}_{pq}$ and their symmetrized analogs $\mathcal{D}^{\bar{\kappa}}_{pq}$ as

$$D^{\bar{\kappa}}_{pq} = D_{pq} + \bar{\kappa}_{pq} \qquad \mathcal{D}^{\bar{\kappa}}_{pq} = D^{\bar{\kappa}}_{pq} + D^{\bar{\kappa}}_{qp} \tag{38}$$

This allows us to derive the final expressions for $\mathscr{F}_{pq}$ and $\mathscr{F}^{\langle d \rangle}_{pq}$:

$$\begin{aligned} \mathscr{F}_{pq} &= \sum_r \left[ \mathcal{D}^{\bar{\kappa}}_{pr} h_{rq} + \sum_J \tilde{\mathcal{W}}^J_{pr} L^J_{qr} \right] + \frac{1}{2} \sum_{rs} \mathcal{D}^{\bar{\kappa}}_{rs} d_{pr} d_{sq} - \frac{\langle d \rangle^2}{2N_e} \mathcal{D}^{\bar{\kappa}}_{pq} + \sqrt{\frac{\omega}{2}} \sum_r \mathcal{D}^{e\text{-}p}_{pr} d_{rq} \\ \mathscr{F}^{\langle d \rangle}_{pi} &= -4 d_{pi} \left( \sum_{rs} D^{\bar{\kappa}}_{rs} d_{rs} - \langle d \rangle - \sqrt{\frac{\omega}{2}} D^p \right) \\ \mathscr{F}^{\langle d \rangle}_{pa} &= 0 \end{aligned} \tag{39}$$

## S4. DENSITY MATRIX EXPRESSIONS

Here, we list programmable expressions for the QED-CCSD-1 density matrices.

### A. One-electron density

Occupied-Occupied block

$$D^e_{ij} = 2\delta_{ij} - \sum_{abk} \bar{t}_{ajbk} t_{aibk} - \sum_{a} \bar{s}_{aj} s_{ai} - \sum_{abk} \bar{s}_{ajbk} s_{aibk} \qquad (40)$$

Virtual-Occupied block

$$D^e_{ai} = \bar{t}_{ai} \qquad (41)$$

Occupied-Virtual block

$$D^e_{ia} = 2\bar{\gamma} s_{ai} + \sum_{bj} \bar{s}_{bj} v_{aibj} + \sum_{bj} \bar{t}_{bj} u_{aibj} - \sum_{bjck} \bar{s}_{bjck} s_{aj} t_{bick} - \sum_{bjck} \bar{s}_{bjck} s_{bi} t_{ajck} \qquad (42)$$

Virtual-Virtual block

$$D^e_{ab} = \sum_{i} \bar{s}_{ai} s_{bi} + \sum_{icj} \bar{s}_{aicj} s_{bicj} + \sum_{icj} \bar{t}_{aicj} t_{bicj} \qquad (43)$$

### B. One-electron-one-photon density

Occupied-Occupied block

$$D^{e\text{-}p}_{ij} = 2\delta_{ij}\bar{\gamma} + 2\delta_{ij}\gamma + 2\sum_{ak} \delta_{ij} \bar{t}_{ak} s_{ak} + \sum_{akbl} \delta_{ij} \bar{t}_{akbl} s_{akbl}$$
$$- \sum_{a} \bar{t}_{aj} s_{ai} - \sum_{a} \bar{s}_{aj} s_{ai} \gamma - \sum_{abk} \bar{s}_{ajbk} t_{aibk} - \sum_{abk} \bar{t}_{ajbk} s_{aibk} - \sum_{abk} \bar{s}_{ajbk} s_{ai} s_{bk} \qquad (44)$$
$$- \sum_{abk} \bar{s}_{ajbk} s_{aibk} \gamma - \sum_{abk} \bar{t}_{ajbk} t_{aibk} \gamma$$

Virtual-Occupied block

$$D^{e\text{-}p}_{ai} = \bar{s}_{ai} + \bar{t}_{ai}\gamma + \sum_{bj} \bar{t}_{aibj} s_{bj} \qquad (45)$$

Occupied-Virtual block

$$D_{ia}^{\text{e-p}} = 2s_{ai} + 2\bar{\gamma}s_{ai}\gamma + \sum_{bj}\bar{s}_{bj}u_{aibj} + \sum_{bj}\bar{t}_{bj}v_{aibj}$$
$$+ 2\sum_{bj}\bar{s}_{bj}s_{ai}s_{bj} - 2\sum_{bj}\bar{s}_{bj}s_{aj}s_{bi} + \sum_{bj}\bar{s}_{bj}v_{aibj}\gamma + \sum_{bj}\bar{t}_{bj}u_{aibj}\gamma$$
$$+ \sum_{bjck}\bar{s}_{bjck}s_{ai}s_{bjck} - 2\sum_{bjck}\bar{s}_{bjck}s_{aj}s_{bick} - 2\sum_{bjck}\bar{s}_{bjck}s_{bi}s_{ajck} + \sum_{bjck}\bar{s}_{bjck}s_{bj}v_{aick} \quad (46)$$
$$- \sum_{bjck}\bar{t}_{bjck}s_{aj}t_{bick} - \sum_{bjck}\bar{t}_{bjck}s_{bi}t_{ajck} + \sum_{bjck}\bar{t}_{bjck}s_{bj}u_{aick}$$
$$- \sum_{bjck}\bar{s}_{bjck}s_{aj}t_{bick}\gamma - \sum_{bjck}\bar{s}_{bjck}s_{bi}t_{ajck}\gamma$$

Virtual-Virtual block

$$D_{ab}^{\text{e-p}} = \sum_{i}\bar{t}_{ai}s_{bi} + \sum_{i}\bar{s}_{ai}s_{bi}\gamma + \sum_{icj}\bar{s}_{aicj}t_{bicj} + \sum_{icj}\bar{t}_{aicj}s_{bicj}$$
$$+ \sum_{icj}\bar{s}_{aicj}s_{bi}s_{cj} + \sum_{icj}\bar{s}_{aicj}s_{bicj}\gamma + \sum_{icj}\bar{t}_{aicj}t_{bicj}\gamma \quad (47)$$

## C. One-photon density

$$D^{p} = \bar{\gamma} + \gamma + \sum_{ak}\bar{t}_{ak}s_{ak} + \frac{1}{2}\sum_{akbl}\bar{t}_{akbl}s_{akbl} \quad (48)$$

## D. Two-electron density

Occupied-Occupied-Occupied-Occupied block

$$d_{ijkl}^{e} = 4\delta_{ij}\delta_{kl} - 2\delta_{il}\delta_{jk} - 2\sum_{a}\delta_{kl}\bar{s}_{aj}s_{ai} + \sum_{a}\delta_{il}\bar{s}_{aj}s_{ak} + \sum_{a}\delta_{jk}\bar{s}_{al}s_{ai} - 2\sum_{a}\delta_{ij}\bar{s}_{al}s_{ak}$$
$$- 2\sum_{abm}\delta_{kl}\bar{s}_{ajbm}s_{aibm} + \sum_{abm}\delta_{il}\bar{s}_{ajbm}s_{akbm} + \sum_{abm}\delta_{jk}\bar{s}_{albm}s_{aibm} - 2\sum_{abm}\delta_{ij}\bar{s}_{albm}s_{akbm}$$
$$- 2\sum_{abm}\delta_{kl}\bar{t}_{ajbm}t_{aibm} + \sum_{abm}\delta_{il}\bar{t}_{ajbm}t_{akbm} + \sum_{abm}\delta_{jk}\bar{t}_{albm}t_{aibm} - 2\sum_{abm}\delta_{ij}\bar{t}_{albm}t_{akbm} \quad (49)$$
$$+ \sum_{ab}\bar{s}_{ajbl}s_{aibk} + \sum_{ab}\bar{t}_{ajbl}t_{aibk}$$

Occupied-Occupied-Occupied-Virtual block

$$
\begin{aligned}
d^e_{ijka} = &-2\delta_{jk}\bar{\gamma}s_{ai} + 4\delta_{ij}\bar{\gamma}s_{ak} \\
&- \sum_{bl}\delta_{jk}\bar{s}_{bl}v_{aibl} + 2\sum_{bl}\delta_{ij}\bar{s}_{bl}v_{akbl} - \sum_{bl}\delta_{jk}\bar{t}_{bl}u_{aibl} + 2\sum_{bl}\delta_{ij}\bar{t}_{bl}u_{akbl} \\
&+ \sum_{blcm}\delta_{jk}\bar{s}_{blcm}s_{al}t_{bicm} - 2\sum_{blcm}\delta_{ij}\bar{s}_{blcm}s_{al}t_{bkcm} \\
&+ \sum_{blcm}\delta_{jk}\bar{s}_{blcm}s_{bi}t_{alcm} - 2\sum_{blcm}\delta_{ij}\bar{s}_{blcm}s_{bk}t_{alcm} \\
&- \sum_{b}\bar{s}_{bj}v_{akbi} - \sum_{b}\bar{t}_{bj}u_{akbi} \\
&+ \sum_{bcl}\bar{s}_{bjcl}s_{ai}t_{bkcl} - 2\sum_{bcl}\bar{s}_{bjcl}s_{ak}t_{bicl} + \sum_{bcl}\bar{s}_{bjcl}s_{al}t_{bick} \\
&+ \sum_{bcl}\bar{s}_{bjcl}s_{bk}t_{aicl} + \sum_{bcl}\bar{s}_{bjcl}s_{ck}t_{albi} - \sum_{bcl}\bar{s}_{bjcl}s_{bi}u_{akcl}
\end{aligned}
\quad (50)
$$

Occupied-Occupied-Virtual-Occupied block

$$
d^e_{ijak} = -\delta_{ik}\bar{t}_{aj} + 2\delta_{ij}\bar{t}_{ak} - \sum_{b}\bar{s}_{akbj}s_{bi} \quad (51)
$$

Occupied-Occupied-Virtual-Virtual block

$$
\begin{aligned}
d^e_{ijab} = &2\sum_{k}\delta_{ij}\bar{s}_{ak}s_{bk} + 2\sum_{kcl}\delta_{ij}\bar{s}_{akcl}s_{bkcl} + 2\sum_{kcl}\delta_{ij}\bar{t}_{akcl}t_{bkcl} \\
&- \bar{s}_{aj}s_{bi} - \sum_{ck}\bar{s}_{ajck}s_{bick} - \sum_{kc}\bar{s}_{akcj}s_{bkci} - \sum_{ck}\bar{t}_{ajck}t_{bick} - \sum_{kc}\bar{t}_{akcj}t_{bkci}
\end{aligned}
\quad (52)
$$

S12

Occupied-Virtual-Occupied-Virtual block

$$d^e_{iajb} = 2u_{aibj} + 2\bar{\gamma}v_{aibj}$$
$$+ 2\sum_{ck}\bar{s}_{ck}s_{ai}u_{bjck} - \sum_{ck}\bar{s}_{ck}s_{aj}u_{bick} - \sum_{ck}\bar{s}_{ck}s_{ak}u_{bjci} - \sum_{ck}\bar{s}_{ck}s_{bi}u_{ajck}$$
$$+ 2\sum_{ck}\bar{s}_{ck}s_{bj}u_{aick} - \sum_{ck}\bar{s}_{ck}s_{bk}u_{aicj} - \sum_{ck}\bar{s}_{ck}s_{ci}u_{akbj} - \sum_{ck}\bar{s}_{ck}s_{cj}u_{aibk}$$
$$+ \sum_{ckdl}\bar{s}_{ckdl}s_{akbl}t_{cidj} + \sum_{ckdl}\bar{s}_{ckdl}s_{akcj}t_{bidl} + \sum_{ckdl}\bar{s}_{ckdl}s_{akdj}t_{blci} + \sum_{ckdl}\bar{s}_{ckdl}s_{bkci}t_{ajdl}$$
$$+ \sum_{ckdl}\bar{s}_{ckdl}s_{bkdi}t_{alcj} + \sum_{ckdl}\bar{s}_{ckdl}s_{cidj}t_{akbl} - \sum_{ckdl}\bar{s}_{ckdl}s_{ajck}u_{bidl} - \sum_{ckdl}\bar{s}_{ckdl}s_{akdl}u_{bjci}$$
$$- \sum_{ckdl}\bar{s}_{ckdl}s_{bick}u_{ajdl} - \sum_{ckdl}\bar{s}_{ckdl}s_{bkdl}u_{aicj} - \sum_{ckdl}\bar{s}_{ckdl}s_{cidl}u_{akbj} - \sum_{ckdl}\bar{s}_{ckdl}s_{cjdl}u_{aibk}$$
$$- \sum_{ckdl}\bar{s}_{ckdl}t_{akdl}v_{bjci} - \sum_{ckdl}\bar{s}_{ckdl}t_{bkdl}v_{aicj} - \sum_{ckdl}\bar{s}_{ckdl}t_{cidl}v_{akbj} - \sum_{ckdl}\bar{s}_{ckdl}t_{cjdl}v_{aibk}$$
$$+ \sum_{ckdl}\bar{s}_{ckdl}u_{aick}v_{bjdl} + \sum_{ckdl}\bar{s}_{ckdl}u_{bjck}v_{aidl} + \sum_{ckdl}\bar{t}_{ckdl}t_{ajck}t_{bldi} + \sum_{ckdl}\bar{t}_{ckdl}t_{akbl}t_{cidj}$$
$$+ \sum_{ckdl}\bar{t}_{ckdl}t_{akdj}t_{blci} - \sum_{ckdl}\bar{t}_{ckdl}t_{akdl}u_{bjci} - \sum_{ckdl}\bar{t}_{ckdl}t_{bick}u_{ajdl} - \sum_{ckdl}\bar{t}_{ckdl}t_{bkdl}u_{aicj}$$
$$- \sum_{ckdl}\bar{t}_{ckdl}t_{cidl}u_{akbj} - \sum_{ckdl}\bar{t}_{ckdl}t_{cjdl}u_{aibk} + \sum_{ckdl}\bar{t}_{ckdl}u_{aick}u_{bjdl} \quad (53)$$

Occupied-Virtual-Virtual-Occupied block

$$d^e_{iabj} = -\sum_k \delta_{ij}\bar{s}_{bk}s_{ak} - \sum_{kcl}\delta_{ij}\bar{s}_{bkcl}s_{akcl} - \sum_{kcl}\delta_{ij}\bar{t}_{bkcl}t_{akcl}$$
$$+ 2\bar{s}_{bj}s_{ai} + \sum_{ck}\bar{s}_{bjck}v_{aick} + \sum_{ck}\bar{t}_{bjck}u_{aick} \quad (54)$$

Occupied-Virtual-Virtual-Virtual block

$$d^e_{iabc} = \sum_j \bar{s}_{bj}v_{aicj} + \sum_j \bar{t}_{bj}u_{aicj}$$
$$+ 2\sum_{jdk}\bar{s}_{bjdk}s_{ai}t_{cjdk} - \sum_{jdk}\bar{s}_{bjdk}s_{aj}t_{cidk} - \sum_{jdk}\bar{s}_{bjdk}s_{ak}t_{cjdi} \quad (55)$$
$$- \sum_{jdk}\bar{s}_{bjdk}s_{ci}t_{ajdk} - \sum_{jdk}\bar{s}_{bjdk}s_{di}t_{akcj} + \sum_{jdk}\bar{s}_{bjdk}s_{cj}u_{aidk}$$

Virtual-Occupied-Virtual-Occupied block

$$d^e_{aibj} = \bar{t}_{aibj} \quad (56)$$

Virtual-Occupied-Virtual-Virtual block

$$d^e_{aibc} = \sum_j \bar{s}_{aibj} s_{cj} \tag{57}$$

Virtual-Virtual-Virtual-Virtual block

$$d^e_{abcd} = \sum_{ij} \bar{s}_{aicj} s_{bidj} + \sum_{ij} \bar{t}_{aicj} t_{bidj} \tag{58}$$


\end{document}